\documentclass[10pt]{revtex4-2}
\usepackage[utf8]{inputenc}
\usepackage{float}
\usepackage{slashed}
\usepackage{wrapfig}
\usepackage{amssymb}
\usepackage{amsmath}
\usepackage{mathtools}
\usepackage{bbold}
\numberwithin{equation}{section}
\usepackage{hyperref}
\usepackage{xcolor}
\DeclareUnicodeCharacter{2113}{l}
\DeclareUnicodeCharacter{2212}{-}

\usepackage[a4paper, top=2.5cm, bottom=2.5cm, left=2cm, right=2cm]{geometry} 
\usepackage[english]{babel} 
\usepackage{flafter} 
\usepackage{multirow}
\usepackage{pifont}
\usepackage{comment}
\usepackage{slashed}
\usepackage[autostyle]{csquotes}
\usepackage{braket}
\usepackage[pdftex]{graphicx} 
\usepackage{makecell}
\usepackage{adjustbox}
\usepackage{caption}
\usepackage{subcaption}
\usepackage{isotope}

\begin{document}

\title{Constraining flavoured leptoquarks with LHC and LFV}

\author{Ivo de Medeiros Varzielas}\email{ivo.de@udo.edu}
\affiliation{CFTP, Departamento de F\'{i}sica, Instituto Superior T\'{e}cnico, Universidade de Lisboa, Avenida Rovisco Pais 1, 1049 Lisboa, Portugal}

\author{Amartya Sengupta}\email{ amartyas@buffalo.edu}\affiliation{CFTP, Departamento de F\'{i}sica, Instituto Superior T\'{e}cnico, Universidade de Lisboa, Avenida Rovisco Pais 1, 1049 Lisboa, Portugal}

\begin{abstract}
     We consider the framework of flavoured leptoquarks, in models with scalar or vector leptoquarks, and look for constraints to the parameter space of these models. Using primarily direct searches at the Large Hadron Collider (LHC) and precision processes that probe Lepton Flavour Violation (LFV), we present lower bounds for the masses of the leptoquarks in different flavoured scenarios. We classify the models according to the specific leptoquark, distinguished with respect to their couplings to charged leptons (lepton isolation, two-columned patterns) and in scenarios with specific hierarchies in the couplings (hierarchical, 2-3 democratic, and flipped).
\end{abstract}

\maketitle
 
\section{Introduction}

The Standard Model (SM) of Particle Physics continues to successfully describe most observations. We know there must be Physics Beyond the Standard Model (BSM) due to e.g. neutrino masses, dark matter, and the baryon asymmetry of the universe.

With the current experimental situation, it is not at all clear what type of BSM extension stands to be the correct description of nature. In recent years, to a great extent due to hints that Lepton Flavour Universality (LFU) might be violated - something which does not occur in the SM - there has been great theoretical interest in extensions including flavoured leptoquarks (LQs), a scenario that has been explored extensively in the literature e.g. 
\cite{deMedeirosVarzielas:2015yxm, Hiller:2016kry,Hiller:2017bzc,  Calibbi:2017qbu,  Hiller:2018wbv, deMedeirosVarzielas:2018bcy,DeMedeirosVarzielas:2019nob,Cornella:2019hct, Bernigaud:2019bfy,Babu:2020hun, Crivellin:2020mjs, Hiller:2021pul,Bernigaud:2021fwn}. 

In December 2022, LHCb released results which supersede the previous results \cite{LHCb:2022zom}. The recent results for $R_{K}$ and $R_{K^{*}}$ are in agreement with the SM prediction.

We summarise the experimental results in Table \ref{tab:bslldata}.
\begin{table}[H]
\begin{center}
\renewcommand{\arraystretch}{1.2}
\begin{tabular}{|c|c|}
     \hline
     Observable & Experiment \\
     \hline
     $R_{K_{[0.1,1.1]}}$ & $0.994\;^{+0.090}_{-0.082}\;({\rm stat})\;^{+0.029}_{-0.027}\;({\rm syst})$ [2022]~\cite{LHCb:2022qnv, LHCb:2022zom} \\
     $R_{{K^{*}}_{[0.1,1.1]}}$ & $0.927\;^{+0.093}_{-0.087}\;({\rm stat})\;^{+0.036}_{-0.035}\;({\rm syst})$ [2022]~\cite{LHCb:2022qnv,LHCb:2022zom} \\
     $R_{K_{[1.1,6]}}$ & $0.949\;^{+0.042}_{-0.041}\;({\rm stat})\;^{+0.022}_{-0.022}\;({\rm syst})$
     [2022]~\cite{LHCb:2022qnv,LHCb:2022zom} \\ 
      $R_{{K^{*}}_{[1.1,6]}}$ & $1.027\;^{+0.072}_{-0.068}\;({\rm stat})\;^{+0.027}_{-0.026}\;({\rm syst})$ 
     [2022]~\cite{LHCb:2022qnv,LHCb:2022zom} \\
      \hline
      $R_{K^*}^{[0.045,1.1]}$ &     $0.66^{+0.11}_{-0.07}\pm0.03$ [2021]~\cite{LHCb:2017avl} \\[5pt]  
      $R_{K^*}^{[1.1,6.0]} $   &     $0.69^{+0.11}_{-0.07}\pm0.05$ [2021]~\cite{LHCb:2017avl} \\[5pt] 
     $R_{K^{\phantom{*}}}^{[1.1,6.0]}$  &    $0.846^{+0.042+0.013}_{-0.039-0.012}$ [2021]~\cite{LHCb:2021trn} \\[2pt]
     \hline
\end{tabular}
\end{center}
\caption{A summary of experimental results for $R_{K}$ and $R_{K^{*}}$. Compare with the SM predictions $R_{K} = 1.00\pm0.01$, $R_{K^*}^{[0.045,1.1]} = 0.906\pm0.028$~\cite{Bordone:2016gaq}, $R_{K^*}^{[1.1,6.0]}=1.00\pm0.01$.}
\label{tab:bslldata}
\end{table} 

This affects the motivation for this type of extension. Nevertheless, models with Flavoured LQs remain interesting and provide a better global fit to flavour data than the SM, see \cite{Allanach:2022iod} with fits using data before and after December 2022. \cite{Ciuchini:2022wbq} considers in detail the impact of the hadronic uncertainties. Another consideration is that CP violation can suppress new physics in the branching rations, as pointed out previously in \cite{Hiller:2014ula} and more recently in \cite{Fleischer:2023zeo} - such that LFU may be violated even with $R_{K}$ and $R_{K^{*}}$ in agreement with their SM values.

Regardless of theoretical bias for flavoured LQs, given current experimental results, it is possible to constrain the parameter space of these models. Some bounds were obtained in \cite{Babu:2020hun} and, in a recent work \cite{Desai:2023jxh}, mass bounds were obtained for LQs coupling either just to electrons or just to muons and democratically to quarks. This is done by analysing the dielectron and dimuon decay channels. 
In this work, we consider several flavoured leptoquark (LQ) scenarios with scalar or vector LQs, with 3 benchmarks of couplings to quarks. We do not consider in detail the flavour patterns where the leptoquark couples only to one letpon flavour, as that is presented in \cite{Hiller:2021pul}. We use Lepton Flavor Violation (LFV) results to provide bounds on the mass of the LQs whenever they apply. We constrain the masses also with results from the Large Hadron Collider (LHC), analysing decay channels with charged leptons and also with neutrinos.

There are complementary bounds to our analysis \cite{Faroughy:2016osc, Allwicher:2022mcg, Greljo:2022jac} arising from t-channel Drell-Yan processes. These will be commented on briefly for comparison with our obtained bounds.

We do not approach here the recent search for
leptoquarks produced resonantly in quark-lepton fusion due to lepton PDFs  \cite{CMS:2023bdh, Buonocore:2020erb, Buonocore:2022msy}. These searches are expected to be relevant for large couplings and large masses \cite{Buonocore:2022msy}, and should be addressed in future work. They are however beyond the our scope here (we note that the reference from 2023 appeared after the present paper).


The layout of this paper is as follows. In Section \ref{sec:models} we review briefly the leptoquark models we will consider. In Section \ref{sec:patterns} we present the flavoured patterns. These are then constrained through experimental results in Section \ref{sec:results}. We conclude in Section \ref{sec:con}.

\section{Leptoquark Models \label{sec:models}}

We consider the following classes of leptoquark models, which we label according to the existing convention as models with the $S_3$, $R_2$, $\Tilde{R}_2$, $\Tilde{S}_1$, $S_1$ and $U_1$ (vector) LQs.

\subsection{Scalar Leptoquarks}

\subsubsection{$S_{3} \sim (\overline{\mathbf{3}},\mathbf{3},1/3)$}

We start by considering the $S_3$ LQ, with SM assignments $(\bar 3,3,1/3)$. This corresponds to an $SU(2)$ triplet that has hypercharge $Y=1/3$. In models with $S_3$, the couplings are
\begin{equation}\label{S3Lagrangian}
    \mathcal{L}_{S_{3}} = \lambda^{ij}_{L}\bar{Q}^{C}_{i}i\tau_{2}(\tau_{k}S^{k}_{3})L_{j} + h.c. \,.
\end{equation}
In this, we have denoted the 3 Pauli matrices as $\tau_{k}$, and the different $SU(2)$ components of $S_3$ are likewise denoted with a superscript. The (Yukawa) couplings of $S_3$ to the fermions are stored in the matrix $\lambda_{L}$.
If $S_3$ couples to quark-quark it can mediate too fast proton decay, so we implicitly assume they vanish or are sufficiently suppressed by some unspecified mechanism (such as the residual symmetries invoked in \cite{deMedeirosVarzielas:2019lgb, Bernigaud:2020wvn}). It is conventional to rewrite the couplings to $SU(2)$ components of $S_3$ with superscripts that identify the respective charge, as follows:
\begin{align}\nonumber
    \mathcal{L}_{S_{3}} &= -\lambda^{ij}_{L} \bar{d^{C}_{L_{i}}}\nu_{L_{j}}S_{3}^{(1/3)} - \sqrt{2}\lambda_{L}^{ij}\bar{d^{C}_{L_{i}}}l_{L_{j}}S_{3}^{(4/3)}\\
    &+ \sqrt{2}(V^{*}\lambda_{L})^{ij}\bar{u^{C}_{L_{i}}}\nu_{L_{j}}S_{3}^{(-2/3)} 
    - (V^{*}\lambda_{L})^{ij}\bar{u^{C}_{L_{i}}}l_{L_{j}}S_{3}^{(1/3)} + h.c. \,.
\end{align}

Similar conventions will be used in the following sections.

\subsubsection{$R_{2} \sim (\mathbf{3},\mathbf{2},7/6)$}

We consider another scalar LQ, $R_2$. In this class of models, the leptoquark has SM assignments $(3,2,7/6)$. This corresponds to an $SU(2)$ doublet that has hypercharge $Y=7/6$.
In models with $R_2$, the couplings are
\begin{equation}\label{R2Lagrangian}
    \mathcal{L}_{R_2} = \lambda^{ij}_{R}\bar{Q}_{i}l_{R_{j}}R_{2} - \lambda^{ij}_{L}\bar{u}_{R_{i}}R_{2}i\tau_{2}L_{j} + h.c.,
\end{equation}
In this case we have (Yukawa) couplings to fermions stored in matrices $\lambda_{L}$ and $\lambda_{R}$. Rewriting with superscripts that identify the respective charge, the terms are as follows
\begin{align}
    \mathcal{L}_{R_{2}} &= (V \lambda_{R})^{ij}\bar{u}_{L_{i}}l_{R_{j}}R_{2}^{(5/3)} + (\lambda_{R})^{ij}\bar{d}_{L_{i}}l_{R_{j}}R_{2}^{(2/3)}\\
    &+ (\lambda_{L})^{ij}\bar{u}_{R_{i}}\nu_{L_{j}}R_{2}^{(2/3)} - (\lambda_{L})^{ij}\bar{u}_{R_{i}} l_{L_{j}}R_{2}^{(5/3)} + h.c. \,.
\end{align}
For our analysis, we have considered the phenomenology of both $\lambda_{L}$ and $\lambda_{R}$. The decays involving both couplings have been listed in \ref{eq: R2decay}.

\subsubsection{$\tilde{R}_2=(\mathbf{3},\mathbf{2},1/6)$}
\label{R_2_tilda}
The coupling of $\tilde{R}_2$ is
\begin{align}
\label{eq:main_t_R_2}
\mathcal{L} &=  -\lambda_{L}^{ij}\bar{d}_{R_i}\tilde{R}_{2} L_{j} +\textrm{h.c.},
\end{align}
where the $SU(2)$ contraction is implicit.

The components of $\tilde{R}_{2}$ have different charges which appear as superscripts below
\begin{align}
\mathcal{L} &= - \lambda_{L}^{ij}\bar{d}_{R_i}e_{L_j}\tilde{R}_{2}^{2/3}+(\lambda_L U)^{ij}\bar{d}_{R_i}\nu_{L_j}\tilde{R}_{2}^{-1/3}+\textrm{h.c.}.
\end{align}
The relation of the couplings to the different charge components is related to the previous couplings through $U$, the PMNS matrix.

\subsubsection{$\tilde{S}_1=(\overline{\mathbf{3}},\mathbf{1},4/3)$}

We employ the same assumption that we followed for $S_3$ and, due to stringent bounds on proton decay, assume the quark-quark coupling of this LQ vanish or is sufficiently suppressed by some unspecified mechanism. In that case, the LQ $\tilde{S}_1$ only has terms
\begin{align}
\label{eq:main_t_S_1}
\mathcal{L} &= + \lambda_{R}^{ij}\bar{d}_{R_i}^{C} \tilde{S}_{1} e_{R_j}+\textrm{h.c.}.
\end{align}
This LQ only couples to right-handed fermions.

\subsubsection{$S_1=(\overline{\mathbf{3}},\mathbf{1},1/3)$}

Again assuming the quark-coupling of this LQ vanishes or is sufficiently suppressed, the LQ $S_1$ has terms
\begin{align}
\mathcal{L} = &+\lambda_{L}^{ij}\bar{Q_i}^{C} S_{1} L_{j}+\lambda_{R_{ij}} \bar{u}_{R_i}^{C} S_{1} e_{R_j}+\textrm{h.c.},
\label{eq:main_S_1}
\end{align}
leading to
\begin{align}
\mathcal{L} = &-(\lambda_{L} U)^{ij} \bar{d}_{L_i}^{C} S_{1} \nu_{L_j}+(V^T \lambda_{L})^{ij}\bar{u}_{L_i}^{C} S_{1} e_{L_j}+\lambda_{R}^{ij}\bar{u}_{R_i}^{C} S_{1} e_{R_j}+\textrm{h.c.}.
\end{align}
The relation of the couplings to the different charge components is related to the previous couplings through $U$, the PMNS matrix, and $V$, the CKM matrix.

\subsection{Vector Leptoquarks}

\subsubsection{$U_{1} \sim (\mathbf{3},\mathbf{1},2/3)$}

We consider also a class of models with vector LQ $U_1$. The SM assignments are $(3,1,2/3)$. This corresponds to an $SU(2)$ singlet. In models with $U_1$, the couplings are
\begin{align}\label{eq:LQLag}
\begin{aligned}
\mathcal{L}_U=&-\frac{1}{2}\,U_{1\,\mu\nu}^\dagger\, U_1^{\mu\nu}+M_{U}^2\,U_{1\,\mu}^\dagger\, U_1^{\mu} -ig_c(1-\kappa_c)\,U_{1\,\mu}^\dagger\,T^a\,U_{1\,\nu}\,G^{a\,\mu\nu} \\
&-i \frac{2}{3} g_{Y}  (1-\kappa_Y)\,U_{1\,\mu}^\dagger\,\,U_{1\,\nu}\,B^{\mu\nu}+(U_1^\mu J_\mu + \mathrm{h.c.}) \,.
\end{aligned}
\end{align}
Following existing conventions in the literature, the covariant derivatives $D_{\mu} = \partial_{\mu} - i g_c\, G_{\mu}^{a}T^{a} - i \frac23 g_{Y}  B_{\mu}$ and $U_{1\, \mu \nu} = D_{\mu} U_{1\,\nu} - D_{\nu} U_{1\,\mu}$.
$T^a$ are $SU(3)_c$ generators, appearing with the ($SU(3)_c$ gluons) $G_\mu^a$ ($a=1,\dots,8$). The ($U(1)_Y$) $B_\mu$ is the SM gauge boson associated with the hypercharge. They are matched with the respective gauge couplings $g_c$ and $g_Y$.

In this class of models, there are parameters which distinguish the origin of the vector leptoquark. $\kappa_c=\kappa_Y=0$ corresponds to the LQ arising from a gauge origin. This should not be the case if $U_1$ arises from strongly-coupled origins.
We can write
\begin{align}
J_\mu=\frac{g_U}{\sqrt{2}}  \left[ \lambda_{L}^{i\alpha}\,(\bar q_{L}^{\,i} \gamma_{\mu}  \ell_{L}^{\alpha})  +     \lambda_{R}^{i \alpha }\,(\bar d_{R}^{\,i}\gamma_{\mu}   e_{R}^{\alpha})\right] \,.
\end{align} 
The $\lambda_L$ and $\lambda_R$ are couplings of the vector LQs to the fermions. We are working on the basis where down-type quarks and charged-leptons have diagonal mass matrices such that we write
\begin{align}\label{eq:DownBasis}
q_L^i=
\begin{pmatrix}
V_{ji}^*\,u^j_L\\
d_L^i
\end{pmatrix}
\,,\qquad\qquad 
\ell_L^i=
\begin{pmatrix}
\nu^i_L\\
e_L^i
\end{pmatrix}
\,.
\end{align}

For the $U_1$ model, we have considered the phenomenology of the scenario $\lambda_{L} = \lambda_{R}$ which will produce the strongest bounds on the LQs. The decays involving this scenario have been listed in \ref{eq: U1decay}.

\subsection{Three benchmarks scenarios}\label{Three scenarios}

In order to obtain indicative mass bounds, we consider the relative strengths of the LQ to the various fermions. In this section, we show the general couplings for all three scenarios. The notation of $\lambda_0$ is generic and should be adapted depending on the specific LQ considered, e.g. in the $R_2$ model, we have both $\lambda_L$ and $\lambda_R$ couplings which are taken into account separately when we use them for the phenomenological analysis. Here, we have considered three benchmark scenarios, \cite{Hiller:2021pul}


\begin{description}
	\item[Hierarchical scenario]	
		The first scenario we consider is the same as in \cite{Hiller:2018wbv} based on flavor models discussed in \cite{Hiller:2016kry}, where we have made the assumption that the hierarchy found in the masses and mixtures in the SM also exists in LQ couplings.
		This would naturally arise in classes of flavour models, including very simple models using the Froggatt-Nielsen mechanism \cite{Froggatt:1978nt} to explain the hierarchies of fermion masses.
In the Frogatt-Nielsen mechanism, an additional (Abelian) symmetry distinguishes the generations of the fermions, forbidding renormalisable Yukawa couplings for the lighter generations. They are allowed at tree level but at a non-renormalizable level through the coupling to an additional scalar field that breaks this additional symmetry. Given that the lighter generations are charged under this symmetry in this scenario, the leptoquark couplings to these fermions are suppressed similarly to the Yukawa couplings. Therefore we consider:
		\begin{equation}
			\lambda_{d\ell} \ :\ \lambda_{s\ell} \ :\ \lambda_{b\ell} \quad\sim\quad \epsilon^3\dots\epsilon^4 \ :\ \epsilon^2 \ :\ 1
		\end{equation}
		between the different quark generations, where $\epsilon\sim 0.2$ is of the order of the Wolfenstein parameter, i.e. the sine of the Cabibbo angle.
		Specifically, we implement
		\begin{equation}
			\lambda_{q l} \sim \lambda_0 \begin{pmatrix} 0 & 0 & 0 \\ \epsilon^2 & \epsilon^2 & \epsilon^2 \\ 1 & 1 & 1 \end{pmatrix}\,.
			\label{eq:scenario_A}
		\end{equation}
   Where, $\lambda_{ql}$ and $\lambda_{0}$ are generic nomenclature for the LQ couplings to quark and leptons. From the eq. \eqref{eq:scenario_A} we can see the coupling is larger for the third generation quarks, therefore with this benchmark the leptoquark decays to $b$ quark and $t$ quark for the down-type and up-type cases respectively.
		We have placed vanishing entries for the first row, experimentally they need to be small enough to respect constraints on rare kaon decays. In cases where the LQ couples to both $\mu$ and $e$ with the same quark, $\mu \to e$ processes come into effect and strongly constrain the couplings, as will be discussed subsequently.

	\item[2-3 Democratic scenario]
		Lastly, we consider a scenario where the couplings to the second and third quark generations are of equal size:
		\begin{equation}
			\lambda_{q l} \sim \lambda_0 \begin{pmatrix} 0 & 0 & 0 \\ 1 & 1 & 1 \\ 1 & 1 & 1 \end{pmatrix}\,.
			\label{eq:scenario_C}
		\end{equation}
  We refer to this as the 2-3 Democratic scenario. In this case, because the coupling is equal for second and third-generation quarks, the LQ decays in equal fractions to second and third-generation quarks for both down and up-type cases. 

  \item[Flipped scenario]
		The second scenario is the inverted hierarchical case,  i.e.:
		\begin{equation}
			\lambda_{q l} \sim \lambda_0 \begin{pmatrix} 0 & 0 & 0 \\ 1 & 1 & 1 \\ \epsilon^2 & \epsilon^2 & \epsilon^2 \end{pmatrix}\,.
			\label{eq:scenario_B}
		\end{equation}
  We refer to this as the flipped scenario. This scenario is opposite to the hierarchical case and with this pattern, the LQ dominantly decays to the $s$ quark for the down-type case and $c$ quark for the up-type case due to the larger coupling constant associated with it.
\end{description}

In all benchmarks we consider, the scalar or vector, the number of parameters is small. We use three for scalar LQs:  the mass, the dominant couplings $\lambda_{b\,l}$ and $\lambda_{s\,l}$. For vector LQs we have these three and also the parameter $\kappa$.
These parameters can be constrained from the measurements of the single- or pair-production cross-section, the corresponding branching fractions and the resonance width, together with the reconstruction of the mass peak.

\section{Testing Flavoured Leptoquarks \label{sec:results}}

In order to place experimental bounds on these flavoured LQs we have to consider different processes and experimental searches. A clear example arises when comparing lepton flavour isolation patterns with the two flavour patterns, where the two flavour patterns can be strongly constrained by Lepton Flavour Violating processes such as $\mu \to e \gamma$ (for $e \mu$), and for LFV meson decays such as $B \to K  \mu^\pm e^\mp$ (and other lepton generations). We don't consider lepton flavour isolation in detail here as those patterns are investigated in \cite{Hiller:2021pul}.

To get mass bounds on the LQs we have studied LQ single and  pair production in $pp$-collisions and LQss decaying to up and down type quarks and different leptons using available search results from the ATLAS and CMS experiments in recent years.
\subsection*{Single and Pair Leptoquark Production and its Decay}\label{Decay}
In this paper, we have considered two dominant mechanisms of leptoquark production at pp colliders. Firstly, the single production of LQs is associated with a lepton and the pair production of LQs.

 For the single production of LQ, we have considered the process $p\,p \to \tau lq$ corresponding to the couplings $\lambda_L$ and $\lambda_R$ for the scalar models and $\lambda_L = \lambda_R$ for the vector model because this is the best set of experimental data available \cite{CMS:2020wzx} with our benchmarks for a single LQ production channel. Since for this type of process, the cross-section is directly proportional to the square of the magnitude of the coupling constant (at the parton level), our chosen three benchmarks very largely affect the production cross-sections which have been shown in the section \ref{SingleProduction}.

Pair production of the LQs is primarily dominated by the strong interaction processes and it doesn't depend on the flavour structure. The flavour structure is mainly responsible for determining the branching fractions into the final products. Based on our three flavour benchmarks  (\ref{eq:scenario_A}), (\ref{eq:scenario_B}), (\ref{eq:scenario_C}) we can experimentally differentiate different patterns of the final products of the LQs in the two-body decays. We hereby list the dominant decays modes for $S_3$, $S_1$, $\Tilde{S}_1$, $R_2$,  $\Tilde{R}_2$,and $U_1$ LQs  \cite{Hiller:2017bzc}\cite{Hiller:2021pul},
\newline For the $S_3$ model the decay modes are
 \begin{eqnarray}\label{Decays}
S_3^{+2/3} &\to& t\ \nu_l \ , \ c\ \nu_l \nonumber \\
S_3^{-1/3} &\to& b \ \nu_l \ ,\ s \ \nu_l \ , \ t\ l^- \ , \  c\ l^-      \label{eq:S3decay}\\
S_3^{-4/3} &\to& b\ l^-, \ s\ l^- \nonumber 
\end{eqnarray}
\newline For the $S_1$ model the decay modes are
 \begin{eqnarray}\label{Decays}
S_1^{-1/3} &\to& b \ \nu_l \ ,\ s \ \nu_l \ , \ t\ l^- \ , \  c\ l^-     
\end{eqnarray}
\newline For the $\Tilde{S}_1$ model the decay modes are
 \begin{eqnarray}\label{Decays}
\Tilde{S}_1^{-4/3} &\to& b\ l^- \ , \  s\ l^-     
\end{eqnarray}
\newline For the $R_2$ model the decay modes are
 \begin{eqnarray}
R_2^{+2/3} &\to& t\ \nu_l \ ,\ c\ \nu_l \ , \ b\ l^+ \ , \  s\ l^+ \nonumber \\
R_2^{+5/3} &\to& \ t\ l^+ \ , \  c\ l^+      \label{eq: R2decay}
\end{eqnarray}
\newline For the $\Tilde{R}_2$ model the decay modes are
 \begin{eqnarray}
\Tilde{R}_2^{+2/3} &\to& b\ l^+ \ , \  s\ l^+ \nonumber \\
\Tilde{R}_2^{-1/3} &\to& b \ \Tilde{\nu}_l \ ,\ s \ \Tilde{\nu}_l \      \label{eq: R2decay}
\end{eqnarray}
We note that in the $R_2$ model $\lambda_L$ is responsible for the decay process $R_2^{+2/3} \to t\ \nu_l \ ,\ c\ \nu_l$ and $\lambda_R$ is responsible for the decay process $R_2^{+2/3} \to b\ l^+ \ , \  s\ l^+.$ For the process $R_2^{+5/3} \to \ t\ l^+ \, \  c\ l^+$ both $\lambda_L$ and $\lambda_R$ are responsible but for our analysis, we have only considered the phenomenology due to the coupling $\lambda_R$ so that we can include the CKM induced effect to the up type quarks. This effect is important because the ATLAS experimental data, which we have used here, distinguish between light and heavy jets. Therefore, the phenomenology corresponding to the coupling $\lambda_R$ for this process is the only relevant one for the current sets of experimental data.   
\newline
\newline
For the singlet $U_1$ leptoquark corresponding the coupling $\lambda_L$ = $\lambda_R$, the decay modes are
\begin{equation} \label{eq: U1decay}
	U_1^{+2/3} \to b\,l^+\,,\ t\,\bar\nu_l \,, \ s\,\ l^+\,,\ c\,\bar\nu_l 
\end{equation}
where $l$ represents the leptons of all generations.
\section{Leptoquark Mass Bounds from LHC Searches}
To evaluate the cross-sections of the different production and decay processes we use \texttt{ Madgraph5\_amc@NLO} to generate events \cite{Alwall:2011uj}. The UFO model files we use here were implemented using Feynrules \cite{Alloul:2013bka}. For the $S_3$, $S_1$, $\Tilde{S}_1$, $\Tilde{R}_2$, and $R_2$ models it has been described in  \cite{Dorsner:2018ynv} and for the $U_1$ model, it is described in \cite{Baker:2019sli}\cite{DiLuzio:2018zxy}\cite{Cornella:2021sby}. The Lagrangian for each model, which has been described in \ref{S3Lagrangian}, \ref{R2Lagrangian} and \ref{eq:LQLag} is implemented in Feynrules. As our main objective is to obtain the mass limits using various cuts provided in the literature, we haven't included any uncertainties or detector simulations here. In the following sections, we present the plots of cross-sections (Y-axis) versus LQ mass (X-axis). For each process, we find an intersection point with the experimental line and thus obtain the respective mass bound for our benchmarks. In some cases where data was insufficient, we extrapolated the experimental data. For convenience, the mass bounds obtained are shown also in Tables. As we run the processes directly by putting the relevant decay states from the LQs based on our benchmarks in the \texttt{ Madgraph5\_amc@NLO,} we didn't have to multiply the branching fractions with the cross-sections.
\newline
\newline
We have done three types of analysis, firstly single production of LQs where we run the process $pp \to \tau LQ$ and we got values of cross-sections for different masses of the LQs. We then used the data from $137 fb^{-1}$, 13 TeV pp collisions\cite{CMS:2020wzx} to find the intersection point for our theory runs. In section \ref{SingleProduction} we have put the plots and table for this run. Here, we discussed the process for scalar LQ models $S_3$, $S_1$, $R_2$, $\Tilde{S}_1$ and $\Tilde{R}_2$ and vector LQ models $U_1$ with $\kappa= $ 0 and $\kappa =$ 1.
\newline
\newline
Secondly, we discuss the pair-produced LQs from pp collisions decaying into down (up) type quarks and leptons of all generations. For the down type quarks and leptons channel we have included $S_3$, $R_2$, $\Tilde{R}_2$, and $\Tilde{S}_1$ models and $U_1$ model with $\kappa = 0$ and $\kappa = 1$ corresponding to $S_3^{-4/3}$, $R_2^{+2/3}$, $\Tilde{S}_1^{-4/3}$, $\Tilde{R}_2^{-2/3}$ and $U_1^{+2/3}$ LQs respectively. In the down sector for the electrons and muons case we have used experimental data from 13 TeV ATLAS pp collisions data at 139 $fb^{-1}$\cite{ATLAS:2020dsk} and for the tau lepton case we have used 13 TeV ATLAS 36.1 $fb^{-1}$ data \cite{ATLAS:2019qpq}. For the tau lepton case, experimental data is available only for the $pp \to LQLQ \to b \tau b \tau$ channel, therefore we implemented our benchmarks to this channel only and we found out the bounds for the flipped ratio case was too weak. So, we have excluded the results for the flipped ratio case here.
\newline
\newline
For the decays to up quarks and leptons, we have analyzed the $S_3$, $S_1$ and $R_2$ models corresponding to the $\lambda_L$, $\lambda_L$ and $\lambda_R$ couplings ($S_3^{-1/3}$, $S_1^{-1/3}$ and $R_2^{+5/3}$) respectively. In the $U_1$ model, we don't get any decay final states to up quarks and leptons. In this section, we have included the CKM-induced effect on the up-type quarks to distinguish them from the light down-type quarks. The ATLAS experimental data also consider this effect to distinguish between light and heavy jets. For the values of the CKM matrix elements we have used the UTFIT Collaboration data from \cite{UTfit:2006vpt}.  We have used experimental data from 13 TeV ATLAS pp collisions data at 139 $fb^{-1}, $ \cite{ATLAS:2020dsk} for charm quarks (for the 2-3 Democratic and flipped ratio) and the first two generations of leptons. For the top quark case (hierarchical) we have used recently published results from ATLAS data \cite{ATLAS:2022wgt}. For the tau lepton case we had data available only for the $t \tau t \tau$ channel from the paper \cite{ATLAS:2021oiz}, therefore we used this data to obtain our mass bounds for our three benchmarks, although as it turned out the mass bound from the flipped case was very weak we didn't include it here.
\newline
\newline
Lastly, we also looked into pair-produced LQs from pp collisions decaying into down (up) type quarks and neutrinos of the first two generations. We didn't include the third generation of neutrinos because there was not enough experimental data available for this channel. In Section \ref{Decay}, we listed the decay channels for each of the LQs. Therefore, we have first put the plots and tables of the process $pp \to S_3^{-1/3} S_3^{-1/3} \to q q \nu_l \nu_l$,\;\;$pp \to S_1^{-1/3} S_1^{-1/3} \to q q \nu_l \nu_l$ and \;\,$pp \to \Tilde{R}_2^{1/3} \Tilde{R}_2^{1/3} \to q q \nu_l \nu_l$,  ($l$ represents $e$ or $\mu$ and $q$ represents down-type quarks $b$ and $s$ quark here) in sections \ref{NEDown} and \ref{NMDown} for our benchmarks and then at sections \ref{NEUp} and \ref{NMUp} we have put the decays of $S_3^{2/3}$, and $U_1^{2/3}$ LQs ($U_1$ LQs for $\kappa =0$ and $\kappa = 1$) which are decaying to up-type quarks and the neutrinos. For the $R_2$ model, we can see that this decay process is possible from the $\lambda_L$ coupling for the $R_2^{2/3}$ LQ, and ran the process in Madgraph, finding the mass-limits to be very low, and for this reason have not considered it further in this section. We first checked the mass bounds using \cite{CMS:2018qqq} data from the $36.1 fb^{-1}$ 2018 CMS paper but the bounds were stronger by considering the bounds with 13 TeV $139 fb^{-1}$ ATLAS data\cite{ATLAS:2022wcu}, which was released very recently. As we obtained stronger bounds, we used the ATLAS data for our analysis. 

\subsubsection{\textbf{Mass bounds from single production of LQs and tau leptons}}\label{SingleProduction}

\begin{figure}[H]
\begin{center}
\includegraphics[scale=0.82]{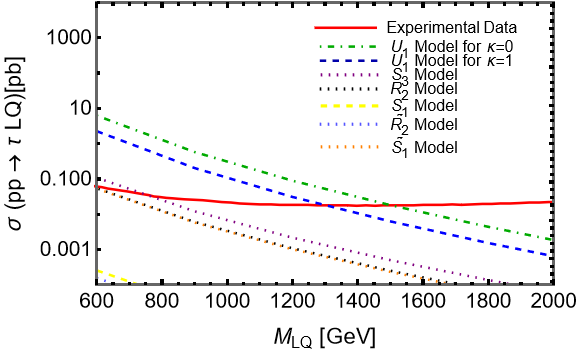}
\includegraphics[scale=0.83]{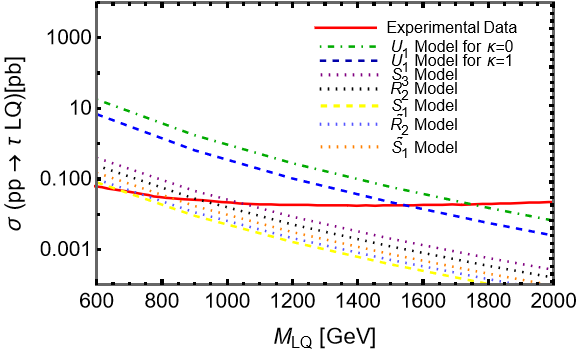}
\includegraphics[scale=0.82]{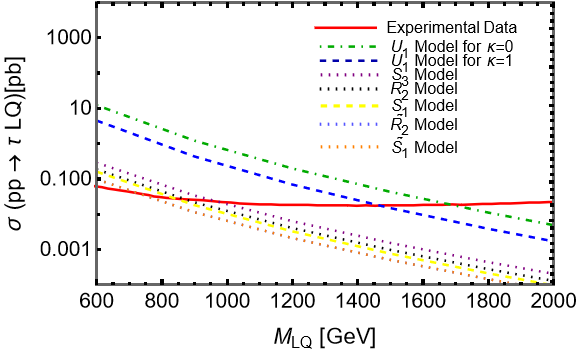}
\end{center}
\caption{Plot for the Cross-section (Y-axis) and LQ Mass(X-axis) for the process $p p \to \tau\,lq$ corresponding to the coupling $\lambda_L$ for the scalar models and $\lambda_L = \lambda_R$ for the vector model.\, a) Upper Left Plot- Hierarchical Scenario \; b) Upper Right Plot- 2-3 Democratic Scenario \; c) Lower Centre Plot- Flipped Scenario\; The solid red line presents the experimental data from 13 TeV, 137 $fb^{-1}$ pp collisions \cite{CMS:2020wzx}. The dotted purple, blue, orange, black, and dashed yellow lines represent our theory results for $S_3$, $\Tilde{R}_2$, $\Tilde{S}_1$, $R_2$ and $S_1$ models respectively, and the dashed blue and dot-dashed green line represent $U_1$ model results for $\kappa = 1$ and $\kappa=0$ respectively. Here, $\lambda_0$ is taken as 1 and $\epsilon$ is $0.2$.  }
\label{fig:PlotLQTauCoupling}
\end{figure}

\begin{table} [H]
\begin{center}
\renewcommand\theadalign{bc}
{\renewcommand{\arraystretch}{0.1}
\begin{tabular}{|c||c|c|c|}
\hline
 LQ Models & Hierarchical  & 2-3 Democratic & Flipped\\
\hline
\hline
\thead{$S_3$ model \\LQ Mass Limit(GeV)} & 752 & 1048 & 972\\
\hline
\thead{$R_2$ model \\LQ Mass Limit(GeV)} & $<600$ & 917 & 872\\
\hline
\thead{$S_1$ model \\LQ Mass Limit(GeV)} & $<600$ & 673 & 841\\
\hline
\thead{$\Tilde{S}_1$ model \\LQ Mass Limit(GeV)} & $<600$ & 831 & 730\\
\hline
\thead{$\Tilde{R}_2$ model \\LQ Mass Limit(GeV)} & $<600$ & 729 & 730\\
\hline
\thead{$U_1$ model \\LQ Mass Limit(GeV) \\ for $\kappa = 0$} & 1503 & 1751 & 1683\\
\hline
\thead{$U_1$ model \\LQ Mass Limit(GeV) \\ for $\kappa = 1$} & 1303 & 1546 & 1468\\
\hline
\end{tabular}}
\caption{Bounds on the LQ mass from single production of LQs and tau leptons.}
\label{tab: Single Production}
\end{center}
\end{table}

\subsubsection{\textbf{Mass Bounds of pair produced LQs decaying into down type quarks and electrons}}

\begin{figure}[H]
\begin{center}
\includegraphics[scale=0.82]{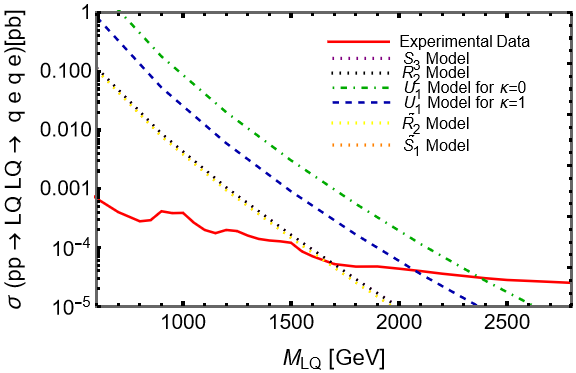}
\includegraphics[scale=0.82]{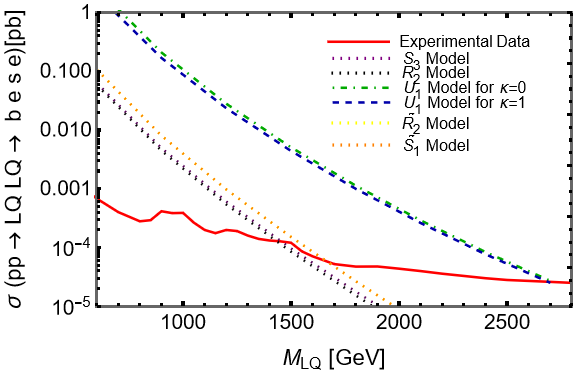}
\includegraphics[scale=0.82]{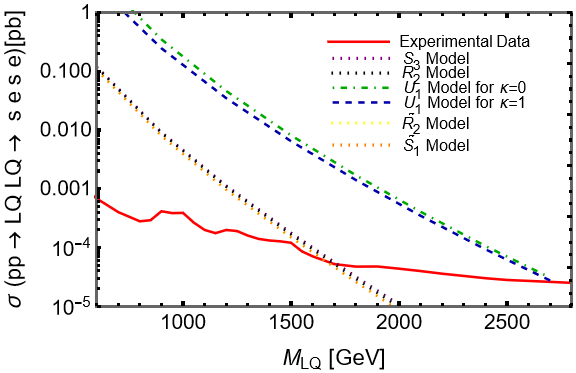}
\end{center}
\caption{Plot for the Cross-section (Y-axis) and LQ Mass(X-axis) for the process $p p \to lq\,lq \to q e^{+} q e^{-},$ where $q$ represents $b$ and $s$ quarks only. Here again, the process is due to $\lambda_L$ coupling for the $S_3$  and $\Tilde{R}_2$ models,\,$\lambda_R$ coupling for the $\Tilde{S}_1$ and $R_2$ models, $\lambda_L = \lambda_R$ for the $U_1$ model.\, a) Upper Left Plot- Hierarchical Scenario. Here $q$ represents $b$ quark only. \; b) Upper Right Plot- 2-3 Democratic Scenario \; c) Lower Centre Plot- Flipped Scenario. The solid red line presents the experimental data from 13 TeV, 139 $fb^{-1}$ pp collisions \cite{ATLAS:2020dsk}. We have used the $q\,e\,q\,e$ plot from the paper because we got better mass bounds. The dotted purple, yellow, orange, and black lines represent our theory results for $S_3$, $\Tilde{R}_2$, $\Tilde{S}_1$ and $R_2$ models respectively and the dashed blue and dot-dashed green line represent $U_1$ model results for $\kappa = 1$ and $\kappa=0$ respectively. We can see that the theory results for $S_3$ and $R_2$ models are very close and we can barely differentiate them. The decay states for each case are different based on the benchmark we have chosen. As we run the process directly in Madgraph, we don't need to multiply the branching fractions with cross-sections. We can also see that the constant $\kappa$ for the $U_1$ model also affects the mass bounds of the LQs based on our benchmarks. Here, $\lambda_0$ is taken as $1$ and $\epsilon$ is $0.2$.}
\label{fig:PlotBEBE}
\end{figure}

\begin{table} [H]
\begin{center}
\renewcommand\theadalign{bc}
{\renewcommand{\arraystretch}{1.2}
\begin{tabular}{|c||c|c|c|}
\hline
 LQ Models & Hierarchical  & 2-3 Democratic & Flipped\\
\hline
\hline
\thead{$S_3$ model \\LQ Mass Limit(GeV)} & 1699 & 1455 & 1718\\
\hline
\thead{$R_2$ model \\LQ Mass Limit(GeV)} & 1695 & 1432 & 1700\\
\hline
\thead{$\Tilde{S}_1$ model \\LQ Mass Limit(GeV)} & 1616 & 1623 & 1619\\
\hline
\thead{$\Tilde{R}_2$ model \\LQ Mass Limit(GeV)} & 1617 & 1622 & 1620\\
\hline
\thead{$U_1$ model \\LQ Mass Limit(GeV) \\ for $\kappa = 0$} & 2374 & 2720 & 2500\\
\hline
\thead{$U_1$ model \\LQ Mass Limit(GeV) \\ for $\kappa = 1$} & 2074 & 2689 & 2453\\
\hline
\end{tabular}}
\caption{Bounds on the LQ mass for pair produced LQs decaying into down-type quarks and electrons.}
\label{tab: DownQuarkElectrons}
\end{center}
\end{table}
\subsubsection{\textbf{Mass Bounds of pair produced LQs decaying into down type quarks and muons}}

\begin{figure}[H]
\begin{center}
\includegraphics[scale=0.82]{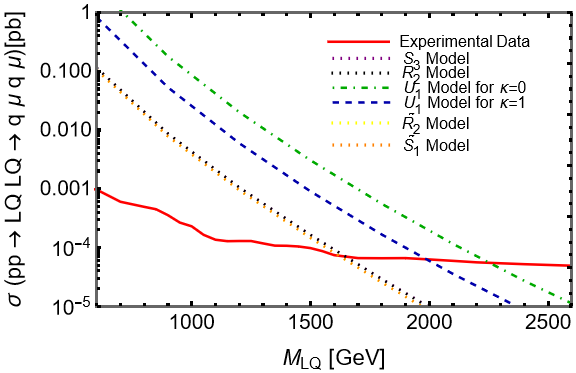}
\includegraphics[scale=0.83]{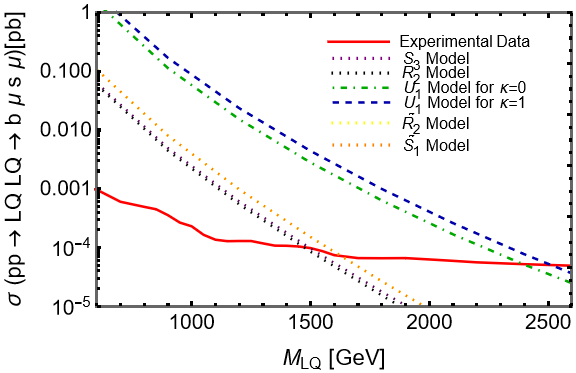}
\includegraphics[scale=0.82]{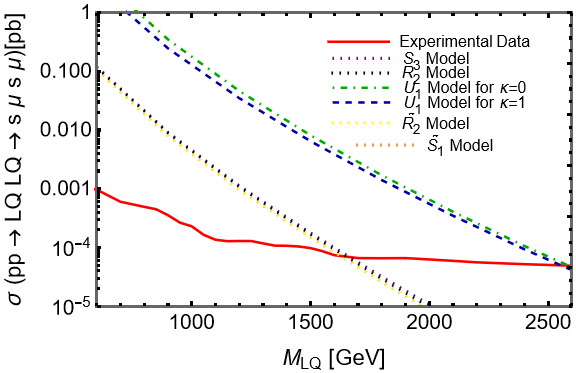}
\end{center}
\caption{Plot for the Cross-section (Y-axis) and LQ Mass(X-axis) for the process $p p \to lq\,lq \to q \mu^{+} q \mu^{-},$  where $q$ represents $b$ and $s$ quarks only. Here, the process is due to $\lambda_L$ coupling for the $S_3$  and $\Tilde{R}_2$ models,$\lambda_R$ coupling for the $\Tilde{S}_1$ and $R_2$ models, $\lambda_L = \lambda_R$ for the $U_1$ model.\, a) Upper Left Plot- Hierarchical Scenario.  Here $q$ represents $b$ quark only. \; b) Upper Right Plot- 2-3 Democratic Scenario \; c) Lower Centre Plot- Flipped Scenario. The solid red line presents the experimental data from 13 TeV, 139 $fb^{-1}$ pp collisions \cite{ATLAS:2020dsk}. We have used the $q\,\mu\, q\,\mu$ plot here from the paper because we got better mass bounds. The dotted purple, yellow, orange, and black lines represent our theory results for $S_3$, $\Tilde{R}_2$, $\Tilde{S}_1$ and $R_2$ models respectively and the dashed blue and dot-dashed green line represent $U_1$ model results for $\kappa = 1$ and $\kappa=0$ respectively. We can see that the theory results for $S_3$ and $R_2$ models are very close and we can barely differentiate them. The decay states for each case are different based on the benchmark we have chosen.  We can again see that the constant $\kappa$ for the $U_1$ model also affects the mass bounds of the LQs based on our benchmarks. Here, $\lambda_0$ is taken as $1$ and $\epsilon$ is $0.2$.}
\label{fig:PlotBMuBMu}
\end{figure}

\begin{table} [H]
\begin{center}
\renewcommand\theadalign{bc}
{\renewcommand{\arraystretch}{1.2}
\begin{tabular}{|c||c|c|c|}
\hline
 LQ Models & Hierarchical  & 2-3 Democratic & Flipped\\
\hline
\hline
\thead{$S_3$ model \\LQ Mass Limit(GeV)} & 1649 & 1497 & 1667\\
\hline
\thead{$R_2$ model \\LQ Mass Limit(GeV)} & 1646 & 1471 & 1649\\
\hline
\thead{$\Tilde{S}_1$ model \\LQ Mass Limit(GeV)} & 1620 & 1623 & 1625\\
\hline
\thead{$\Tilde{R}_2$ model \\LQ Mass Limit(GeV)} & 1620 & 1624 & 1625\\
\hline
\thead{$U_1$ model \\LQ Mass Limit(GeV) \\ for $\kappa = 0$} & 2254 & 2394 & 2725\\
\hline
\thead{$U_1$ model \\LQ Mass Limit(GeV) \\ for $\kappa = 1$} & 1988 & 2513 & 2705\\
\hline
\end{tabular}}
\caption{Bounds on the LQ mass for pair produced LQs decaying into down type quarks and muons.}
\label{tab:DownQuarkMuons}
\end{center}
\end{table}
\subsubsection{\textbf{Mass Bounds of pair produced LQs decaying into down type quarks and tau leptons}}
\begin{figure}[H]
\begin{center}
\includegraphics[scale=0.82]{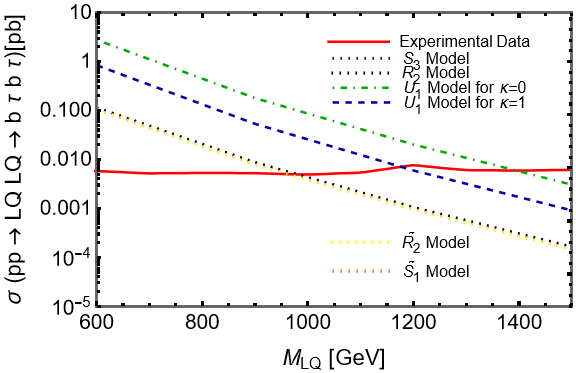}
\includegraphics[scale=0.82]{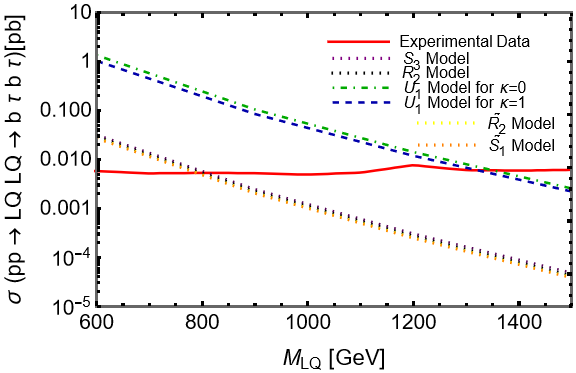}
\end{center}
\caption{Plot for the Cross-section (Y-axis) and LQ Mass(X-axis) for the process $p p \to lq\,lq \to b \tau^{+} b \tau^{-}$. Here again, the process is due to $\lambda_L$ coupling for the $S_3$  and $\Tilde{R}_2$ models,$\lambda_R$ coupling for the $\Tilde{S}_1$ and $R_2$ models, $\lambda_L = \lambda_R$ for the $U_1$ model.\, a) Upper Left Plot- Hierarchical Scenario.  Here $q$ represents $b$ quark only.  \; b) Upper Right Plot- 2-3 Democratic Scenario. \; The solid red line presents the experimental data from 13 TeV, 36.1 $fb^{-1}$ pp collisions \cite{ATLAS:2019qpq}. We have used the $b\, \tau\, b \,\tau$ plot from the paper. The dotted purple, yellow, orange, and black lines represent our theory results for $S_3$, $\Tilde{R}_2$, $\Tilde{S}_1$ and $R_2$ models respectively and the dashed blue and dot-dashed green line represent $U_1$ model results for $\kappa = 1$ and $\kappa=0$ respectively. We can see here that the constant $\kappa$ for the $U_1$ model also affects the mass bounds of the LQs based on our benchmarks. Here, we haven't plotted the results from the flipped ratio scenario because the mass bounds were very weak due to small coupling to the bottom quark.  Here again, $\lambda_0$ is taken as $1$ and $\epsilon$ is $0.2$.}
\label{fig:PlotBTauBTau}
\end{figure}

\begin{table} [H]
\begin{center}
\renewcommand\theadalign{bc}
{\renewcommand{\arraystretch}{1.2}
\begin{tabular}{|c||c|c|}
\hline
 LQ Models & Hierarchical  & 2-3 Democratic\\
\hline
\hline
\thead{$S_3$ model \\LQ Mass Limit(GeV)} & 981 & 810\\
\hline
\thead{$R_2$ model \\LQ Mass Limit(GeV)} & 982 & 800\\
\hline
\thead{$\Tilde{S}_1$ model \\LQ Mass Limit(GeV)} & 964 & 788\\
\hline
\thead{$\Tilde{R}_2$ model \\LQ Mass Limit(GeV)} & 964 & 791\\
\hline
\thead{$U_1$ model \\LQ Mass Limit(GeV) \\ for $\kappa = 0$} & 1395 & 1350\\
\hline
\thead{$U_1$ model \\LQ Mass Limit(GeV) \\ for $\kappa = 1$} & 1177 & 1321\\
\hline
\end{tabular}}
\caption{Bounds on the LQ mass for pair produced LQs decaying into bottom quarks and tau leptons. The Flipped ratio benchmark hasn't been placed here due to very weak limits.}
\label{tab: BQuarkTau}
\end{center}
\end{table}
\newpage
\subsubsection{\textbf{Mass Bounds of pair produced LQs decaying into up type quarks and electrons}}
\begin{figure}[H]
\begin{center}
\includegraphics[scale=0.82]{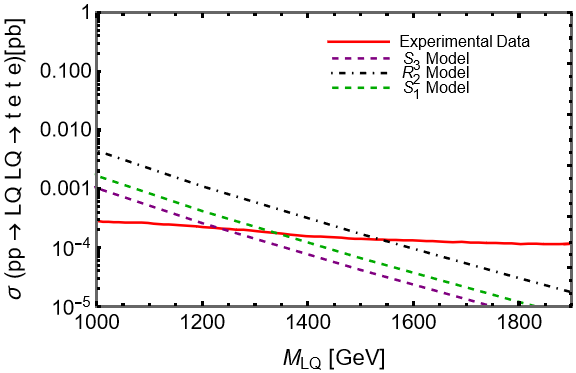}
\includegraphics[scale=0.83]{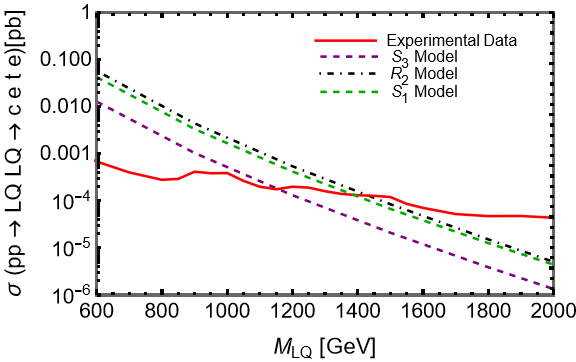}
\includegraphics[scale=0.82]{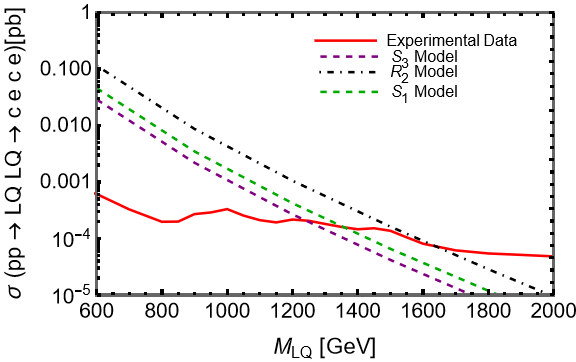}
\end{center}
\caption{Plot for the Cross-section (Y-axis) and LQ Mass(X-axis) for the process $p p \to lq\,lq \to q e^{+} q e^{-},$ where $q$ represents $c$ and $t$ quarks only. Here, the process is due to $\lambda_L$ coupling for the $S_3$, $\lambda_R$ coupling for the $R_2$ model and both $\lambda_L$= $\lambda_R$ couplings for the $S_1$ model.\, a) Upper Left Plot- Hierarchical Scenario \; b) Upper Right Plot- 2-3 Democratic Scenario \; c) Lower Centre Plot- Flipped Scenario. The solid red line presents the experimental data from 13 TeV, 139 $fb^{-1}$ pp collisions \cite{ATLAS:2020dsk} and \cite{ATLAS:2022wgt}. We have used the $q\,e\,q\,e$ plot from the paper \cite{ATLAS:2020dsk} for the 2-3 Democratic and Flipped case and $t\, e \, t \, e$ plot from \cite{ATLAS:2022wgt} paper for the Hierarchical case due to better limits.  The dashed purple, dashed green, and dot-dashed black lines represent our theory results for $S_3$, $S_1$, and $R_2$ models respectively. The decay states for each case are different based on the benchmark we have chosen. As we run the process directly in Madgraph, we don't need to multiply the branching fractions with cross-sections.  Here also, $\lambda_0$ is taken as $1$ and $\epsilon$ is $0.2$.}
\label{fig:PlotTETE}
\end{figure}

\begin{table} [H]
\begin{center}
\renewcommand\theadalign{bc}
{\renewcommand{\arraystretch}{1.2}
\begin{tabular}{|c||c|c|c|}
\hline
 LQ Models & Hierarchical  & 2-3 Democratic & Flipped\\
\hline
\hline
\thead{$S_3$ model \\LQ Mass Limit(GeV)} & 1234 & 1156 & 1241 \\
\hline
\thead{$R_2$ model \\LQ Mass Limit(GeV)} & 1538 & 1432 & 1639\\
\hline
\thead{$S_1$ model \\LQ Mass Limit(GeV)} & 1340 & 1387 & 1359\\
\hline
\end{tabular}}
\caption{Bounds on the LQ mass for pair produced LQs decaying into up-type quarks and electrons.}
\label{tab:UpQuarkElectrons}
\end{center}
\end{table}
\newpage
\subsubsection{\textbf{Mass Bounds of pair produced LQs decaying into up type quarks and muons}}
\begin{figure}[H]
\begin{center}
\includegraphics[scale=0.80]{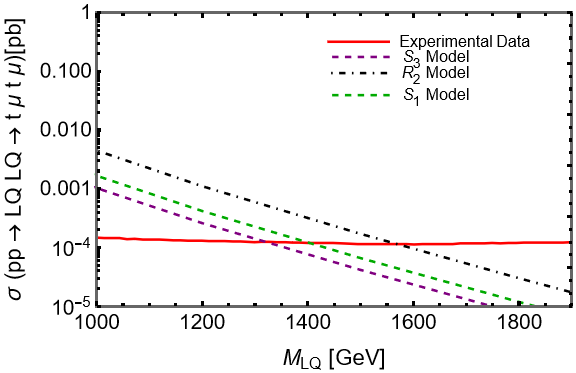}
\includegraphics[scale=0.82]{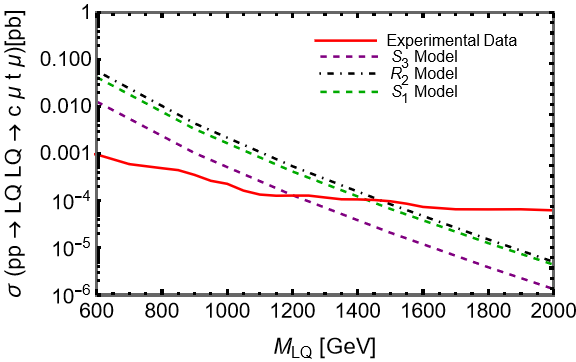}
\includegraphics[scale=0.82]{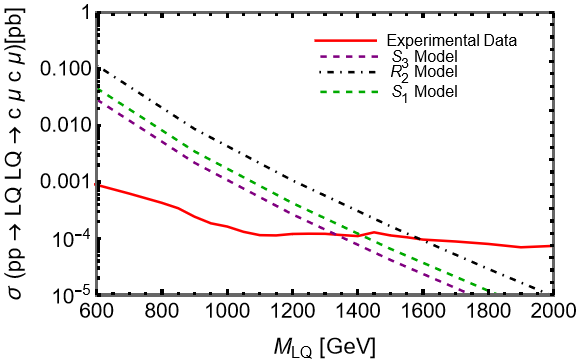}
\end{center}
\caption{Plot for the Cross-section (Y-axis) and LQ Mass(X-axis) for the process $p p \to lq\,lq \to q \mu^{+} q \mu^{-},$ where $q$ represents $c$ and $t$ quarks only. Here, the process is due to $\lambda_L$ coupling for the $S_3$, $\lambda_R$ coupling for the $R_2$ model and both $\lambda_L$ = $\lambda_R$ couplings for the $S_1$ model.\, a) Upper Left Plot- Hierarchical Scenario \; b) Upper Right Plot- 2-3 Democratic Scenario \; c) Lower Centre Plot- Flipped Scenario. The solid red line presents the experimental data from 13 TeV, 139 $fb^{-1}$ pp collisions \cite{ATLAS:2020dsk} and \cite{ATLAS:2022wgt}. We have used the $q\, \mu\, q \,\mu$ plot from the paper \cite{ATLAS:2020dsk} for the 2-3 Democratic and Flipped case and $t \, \mu \, t \, \mu$ plot from \cite{ATLAS:2022wgt} for the Hierarchical case because of better limits.  The dashed purple, dashed green, and dot-dashed black lines represent our theory results for $S_3$, $S_1$, and $R_2$ models respectively. The decay states for each case are different based on the benchmark we have chosen. As we run the process directly in Madgraph, we don't need to multiply the branching fractions with cross-sections.  Here, $\lambda_0$ is taken as $1$ and $\epsilon$ is $0.2$.}
\label{fig:PlotTMuTMu}
\end{figure}

\begin{table} [H]
\begin{center}
\renewcommand\theadalign{bc}
{\renewcommand{\arraystretch}{1.2}
\begin{tabular}{|c||c|c|c|}
\hline
 LQ Models & Hierarchical  & 2-3 Democratic & Flipped\\
\hline
\hline
\thead{$S_3$ model \\LQ Mass Limit(GeV)} & 1320 & 1207 & 1334 \\
\hline
\thead{$R_2$ model \\LQ Mass Limit(GeV)} & 1569 & 1471 & 1591\\
\hline
\thead{$S_1$ model \\LQ Mass Limit(GeV)} & 1404 & 1426 & 1411 \\
\hline
\end{tabular}}
\caption{Bounds on the LQ mass for pair produced LQs decaying into up type quarks and muons.}
\label{tab:UpQuarkMuons}
\end{center}
\end{table}
\newpage
\subsubsection{\textbf{Mass Bounds of pair produced LQs decaying into up type quarks and tau leptons}}
\begin{figure}[H]
\begin{center}
\includegraphics[scale=0.82]{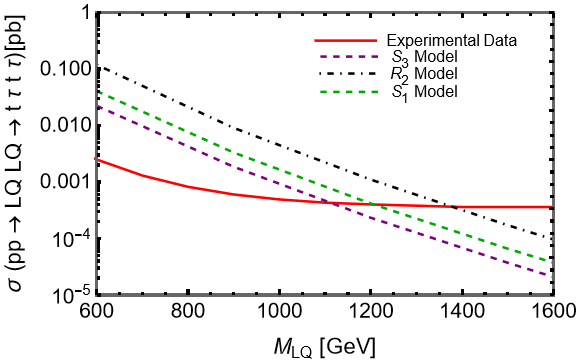}
\includegraphics[scale=0.82]{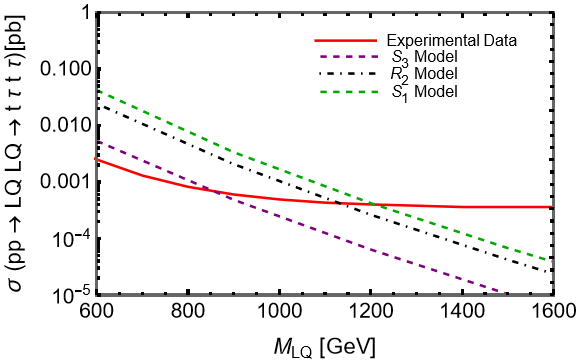}
\end{center}
\caption{Plot for the Cross-section (Y-axis) and LQ Mass(X-axis) for the process $p p \to lq\,lq \to t \tau^{+} t \tau^{-}$.\, Here, the process is due to $\lambda_L$ coupling for the $S_3$, $\lambda_R$ coupling for the $R_2$ model and both $\lambda_L$ = $\lambda_R$ couplings for the $S_1$ model.\, a) Upper Left Plot- Hierarchical Scenario \; b) Upper Right Plot- 2-3 Democratic Scenario.\; The solid red line presents the experimental data from 13 TeV, 139 $fb^{-1}$ pp collisions \cite{ATLAS:2021oiz}. We have used the $t\, \tau\, t \,\tau$ plot from the paper. The dashed purple, dashed green, and dot-dashed black lines represent our theory results for $S_3$, $S_1$, and $R_2$ models respectively. Here, we haven't plotted the results from the flipped ratio scenario because the mass bounds were very weak due to the small coupling to the top quark.  Here, $\lambda_0$ is taken as $1$ and $\epsilon$ is $0.2$.}
\label{fig:PlotTTauTTau}
\end{figure}

\begin{table} [H]
\begin{center}
\renewcommand\theadalign{bc}
{\renewcommand{\arraystretch}{1.2}
\begin{tabular}{|c||c|c|}
\hline
 LQ Models & Hierarchical  & 2-3 Democratic \\
\hline
\hline
\thead{$S_3$ model \\LQ Mass Limit(GeV)} & 1113 & 856 \\
\hline
\thead{$R_2$ model \\LQ Mass Limit(GeV)} & 1378 & 1131 \\
\hline
\thead{$S_1$ model \\LQ Mass Limit(GeV)} & 1206 & 1209 \\
\hline
\end{tabular}}
\caption{Bounds on the LQ mass for pair produced LQs decaying into top quarks and tau leptons.}
\label{tab:TopQuarkTauLepton}
\end{center}
\end{table}
\newpage
\subsubsection{\textbf{Mass Bounds of pair produced LQs decaying into down type quarks and electron neutrinos}}\label{NEDown}
\begin{figure}[H]
\begin{center}
\includegraphics[scale=0.85]{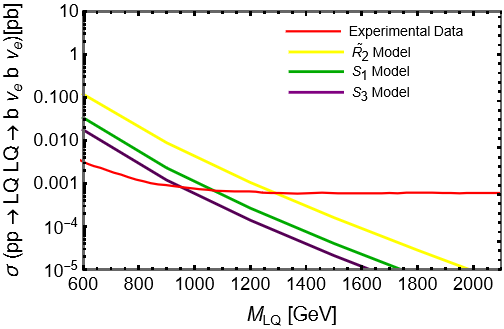}
\includegraphics[scale=0.85]{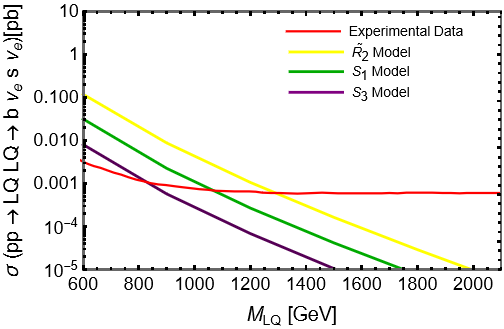}
\includegraphics[scale=0.85]{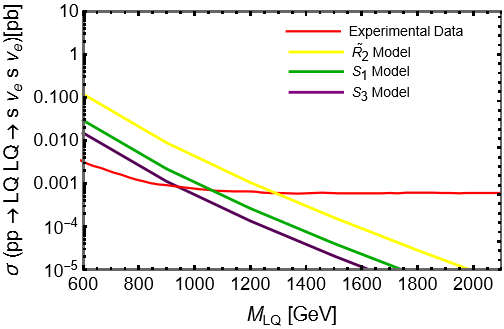}
\end{center}
\caption{Plot for the Cross-section (Y-axis) and LQ Mass(X-axis) for the process $p p \to lq\,lq \to q\, \nu_e \,q \,\nu_e,$  where $q$ represents $b$ and $s$ quarks only. Here, the process is due to $\lambda_L$ coupling for the $S_3$ model.\, a) Upper Left Plot- Hierarchical Scenario \; b) Upper Right Plot- 2-3 Democratic Scenario \; c) Lower Centre Plot- Flipped Scenario.\, The solid red line presents the experimental data from 13 TeV, 139 $fb^{-1}$ pp collisions \cite{ATLAS:2022wcu}. We have used the $q\, \nu_e \,q \,\nu_e$ plot from the paper for this analysis. The purple line represents our theory results for the $S_3$ model, the green line represents the $S_1$ model, and the yellow line represents the $\Tilde{R}_2$ model. We can see the $\Tilde{R}_2$ model gives us the strongest constraint for the LQ mass for this channel. The decay states for each case are different based on the benchmarks we have chosen. As we run the process directly in Madgraph, we don't need to multiply the branching fractions with cross-sections.  Here, $\lambda_0$ is taken as $1$ and $\epsilon$ is $0.2$.}
\label{fig:PlotNeutrinoElectronDown}
\end{figure}
\begin{table} [H]
\begin{center}
\renewcommand\theadalign{bc}
{\renewcommand{\arraystretch}{1.2}
\begin{tabular}{|c||c|c|c|}
\hline
 LQ Models & Hierarchical  & 2-3 Democratic & Flipped\\
\hline
\hline
\thead{$S_3$ model \\LQ Mass Limit(GeV)} & 952 & 828 & 939 \\
\hline
\thead{$S_1$ model \\LQ Mass Limit(GeV)} & 1068 & 1067 & 1062\\
\hline
\thead{$\Tilde{R}_2$ model \\LQ Mass Limit(GeV)} & 1290 & 1291 & 1290\\
\hline
\end{tabular}}
\caption{Bounds on the LQ mass for pair produced LQs decaying into down-type quarks and electron neutrinos.}
\label{tab:DownQElectronNeutrinos}
\end{center}
\end{table}
\newpage
\subsubsection{\textbf{Mass Bounds of pair produced LQs decaying into down type quarks and muon neutrinos}}\label{NMDown}
\begin{figure}[H]
\begin{center}
\includegraphics[scale=0.85]{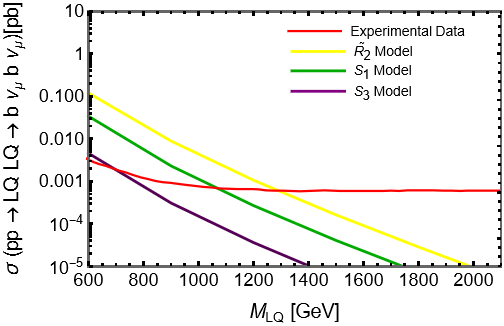}
\includegraphics[scale=0.85]{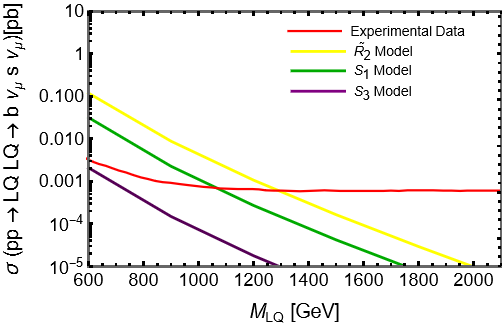}
\includegraphics[scale=0.85]{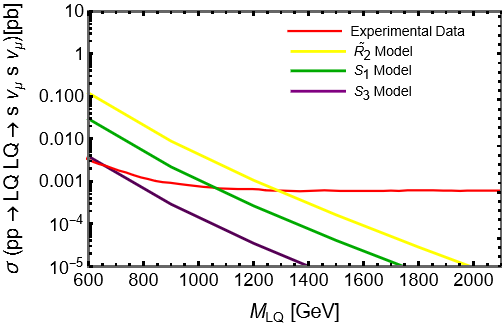}
\end{center}
\caption{Plot for the Cross-section (Y-axis) and LQ Mass(X-axis) for the process $p p \to lq\,lq \to q\, \nu_\mu \,q \,\nu_\mu,$  where $q$ represents $b$ and $s$ quarks only. \, a) Upper Left Plot- Hierarchical Scenario \; b) Upper Right Plot- 2-3 Democratic Scenario \; c) Lower Centre Plot- Flipped Scenario. The solid red line presents the experimental data from 13 TeV, 139 $fb^{-1}$ pp collisions \cite{ATLAS:2022wcu}. We have used the $q \,\nu_\mu \,q \,\nu_\mu$ plot from the paper for this analysis. The purple line represents our theory results for the $S_3$ model, the green line represents the $S_1$ model, and the yellow line represents the $\Tilde{R}_2$ model. We can see the $\Tilde{R}_2$ model gives us the strongest constraint for the LQ mass for this channel. The decay states for each case are different based on the benchmarks we have chosen. As we run the process directly in Madgraph, we don't need to multiply the branching fractions with cross-sections.  Here again, $\lambda_0$ is taken as $1$ and $\epsilon$ is $0.2$.}
\label{fig:PlotNeutrinoMuonDown}
\end{figure}
\begin{table} [H]
\begin{center}
\renewcommand\theadalign{bc}
{\renewcommand{\arraystretch}{1.2}
\begin{tabular}{|c||c|c|c|}
\hline
 LQ Models & Hierarchical  & 2-3 Democratic & Flipped\\
\hline
\hline
\thead{$S_3$ model \\LQ Mass Limit(GeV)} & 695 & 580 & 656 \\
\hline
\thead{$S_1$ model \\LQ Mass Limit(GeV)} & 1069 & 1068 & 1063\\
\hline
\thead{$\Tilde{R}_2$ model \\LQ Mass Limit(GeV)} & 1291 & 1289 & 1291\\
\hline
\end{tabular}}
\caption{Bounds on the LQ mass for pair produced LQs decaying into down-type quarks and muon neutrinos.}
\label{tab:DownQMuonNeutrinos}
\end{center}
\end{table}
\newpage
\subsubsection{\textbf{Mass Bounds of pair produced LQs decaying into up type quarks and electron neutrinos}}\label{NEUp}
\begin{figure}[H]
\begin{center}
\includegraphics[scale=0.84]{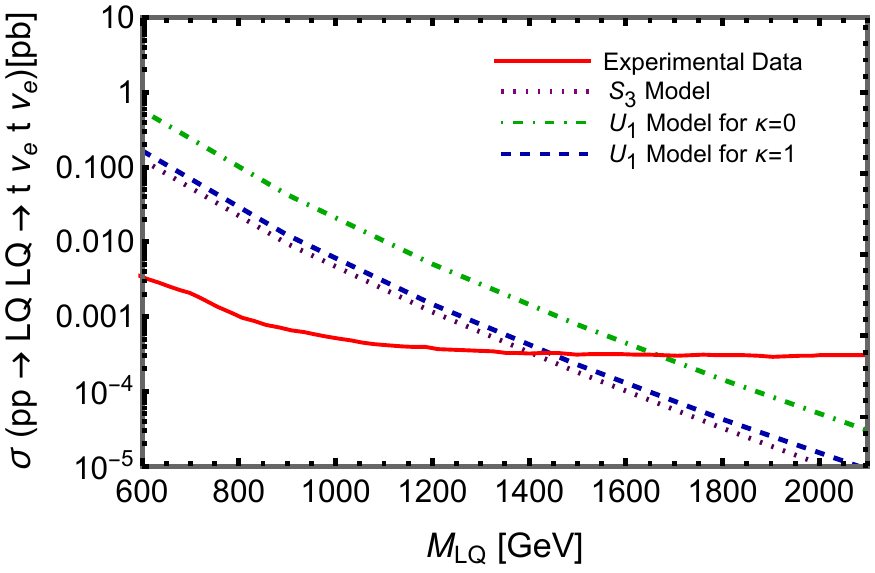}
\includegraphics[scale=0.84]{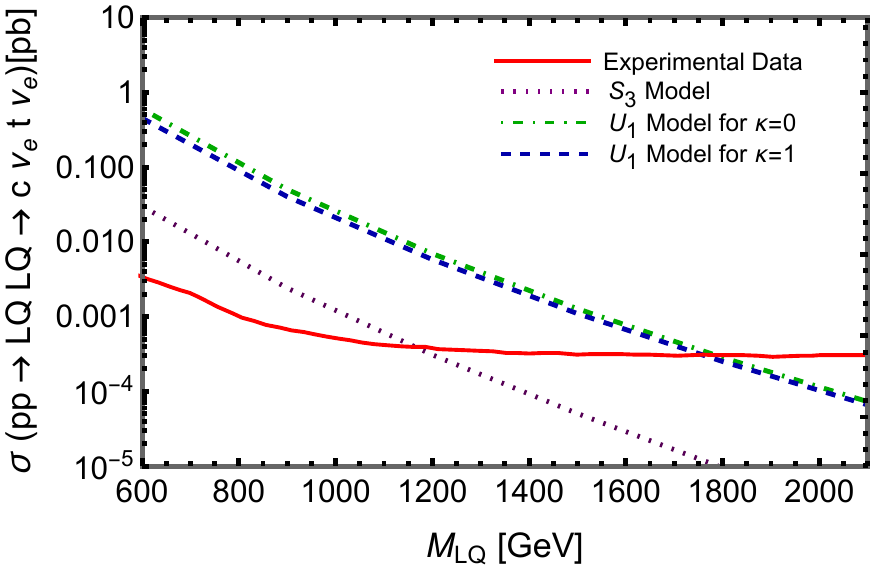}
\includegraphics[scale=0.84]{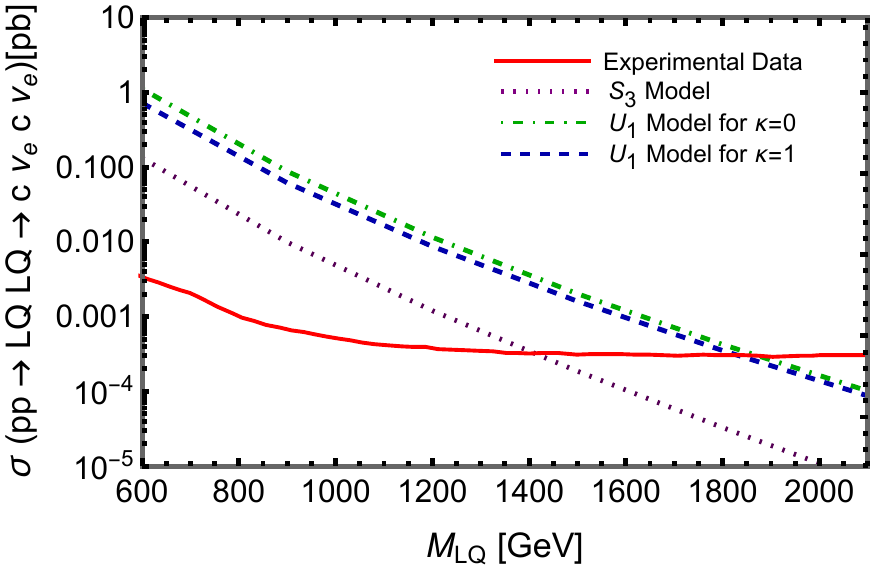}
\end{center}
\caption{Plot for the Cross-section (Y-axis) and LQ Mass(X-axis) for the process $p p \to lq\,lq \to q\, \nu_e \,q \,\nu_e,$ where $q$ represents $c$ and $t$ quarks only.  Here, the process is due to $\lambda_L$ coupling for the $S_3$ model and $\lambda_L = \lambda_R$ for the $U_1$ model. For the $R_2$ model, we can see that this decay process is possible corresponding to $\lambda_L$ coupling, we, therefore, run the process in the Madgraph but we found that the mass-limits are well below $500$ GeV, therefore we haven't plotted it here. a) Upper Left Plot- Hierarchical Scenario \; b) Upper Right Plot- 2-3 Democratic Scenario \; c) Lower Centre Plot- Flipped Scenario. \, The solid red line presents the experimental data from 13 TeV, 139 $fb^{-1}$ pp collisions \cite{ATLAS:2022wcu}. We have used the $q \,\nu_e \,q \,\nu_e$ plot from the paper for this analysis. The dotted purple line represents our theory results for the $S_3$ model, and the dashed blue and dot-dashed green line represent our theory results for the $\kappa = 1$ and $\kappa = 0$ respectively. The decay states for each case are different based on the benchmarks we have chosen. As we run the process directly in Madgraph, we don't need to multiply the branching fractions with cross-sections. We can again see that the constant $\kappa$ for the $U_1$ model also affects the mass bounds of the LQs based on our benchmarks.  Here, $\lambda_0$ is taken as $1$ and $\epsilon$ is $0.2$.}
\label{fig:PlotNeutrinoElectronUP}
\end{figure}
\begin{table} [H]
\begin{center}
\renewcommand\theadalign{bc}
{\renewcommand{\arraystretch}{1.2}
\begin{tabular}{|c||c|c|c|}
\hline
 LQ Models & Hierarchical  & 2-3 Democratic & Flipped\\
\hline
\hline
\thead{$S_3$ model \\LQ Mass Limit(GeV)} & 1406 & 1165 & 1411 \\
\hline
\thead{$U_1$ model \\LQ Mass Limit(GeV) \\ for $\kappa = 0$}  & 1665 & 1785 & 1872\\
\hline
\thead{$U_1$ model \\LQ Mass Limit(GeV) \\ for $\kappa = 1$}  & 1441 & 1757 & 1830\\
\hline
\end{tabular}}
\caption{Bounds on the LQ mass for pair produced LQs decaying into up-type quarks and electron neutrinos.}
\label{tab:NeutrinoUpQuarkElectrons}
\end{center}
\end{table}
\newpage
\subsubsection{\textbf{Mass Bounds of pair produced LQs decaying into up type quarks and muon neutrinos}}\label{NMUp}
\begin{figure}[H]
\begin{center}
\includegraphics[scale=0.84]{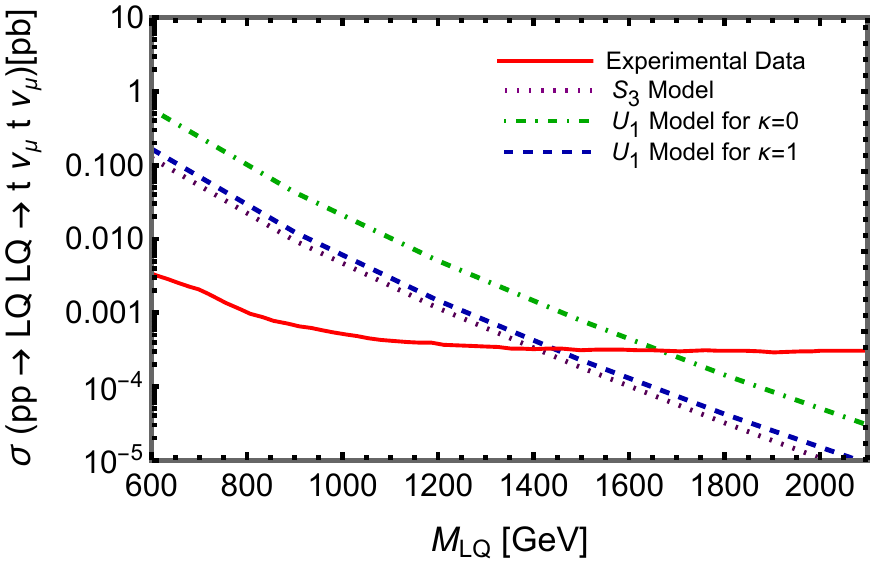}
\includegraphics[scale=0.84]{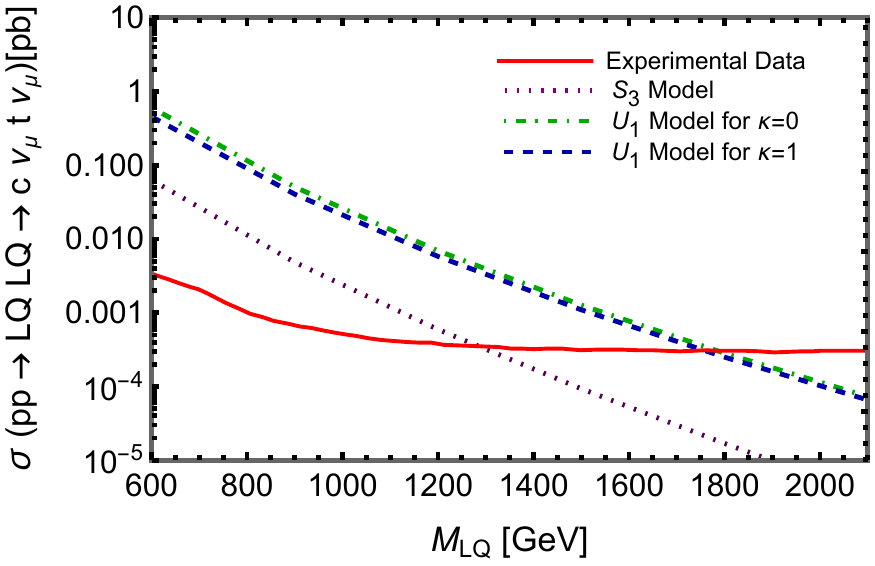}
\includegraphics[scale=0.84]{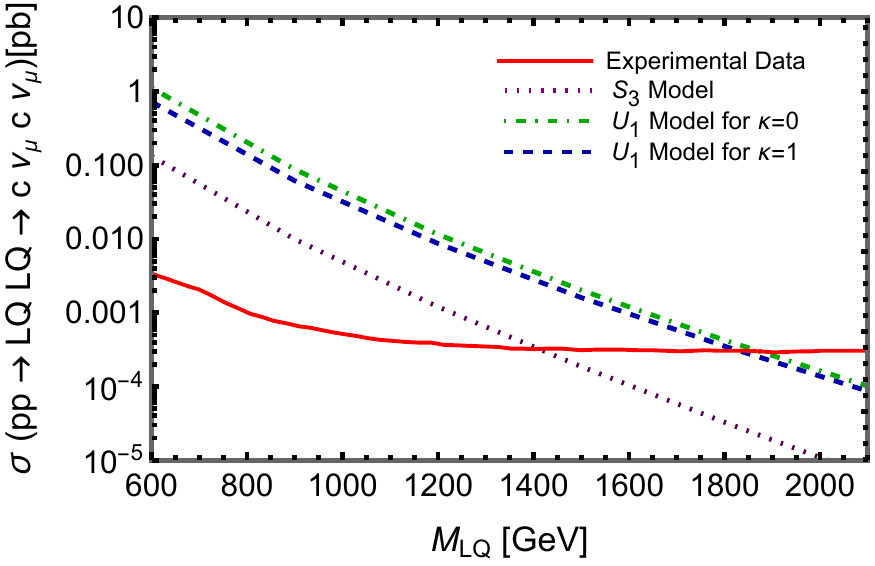}
\end{center}
\caption{Plot for the Cross-section (Y-axis) and LQ Mass(X-axis) for the process $p p \to lq\,lq \to q\, \nu_\mu \,q \,\nu_\mu,$ where $q$ represents $c$ and $t$ quarks only. Here, the process is due to $\lambda_L$ coupling for the $S_3$ model and $\lambda_L = \lambda_R$ for the $U_1$ model.\, For the $R_2$ model we can see that this decay process is possible corresponding to $\lambda_L$ coupling, we run the process in the Madgraph but we found that the mass-limits are well below $500$ GeV, therefore we haven't plotted it here. a) Upper Left Plot- Hierarchical Scenario \; b) Upper Right Plot- 2-3 Democratic Scenario \; c) Lower Centre Plot- Flipped Scenario \, The solid red line presents the experimental data from 13 TeV, 139 $fb^{-1}$ pp collisions \cite{ATLAS:2022wcu}. We have used the $q\, \nu_\mu\, q \,\nu_\mu$ plot from the paper for this analysis. The dotted purple line represents our theory results for the $S_3$ model, and the dashed blue and dot-dashed green lines represent our theory results for the $\kappa = 1$ and $\kappa = 0$ respectively. The decay states for each case are different based on the benchmarks we have chosen. As we run the process directly in Madgraph, we don't need to multiply the branching fractions with cross-sections. We can again see that the constant $\kappa$ for the $U_1$ model also affects the mass bounds of the LQs based on our benchmarks.  Here also, $\lambda_0$ is taken as $1$ and $\epsilon$ is $0.2$.}
\label{fig:PlotNeutrinoMuonUP}
\end{figure}
\begin{table} [H]
\begin{center}
\renewcommand\theadalign{bc}
{\renewcommand{\arraystretch}{1.2}
\begin{tabular}{|c||c|c|c|}
\hline
 LQ Models & Hierarchical  & 2-3 Democratic & Flipped\\
\hline
\hline
\thead{$S_3$ model \\LQ Mass Limit(GeV)} & 1406 & 1165 & 1412 \\
\hline
\thead{$U_1$ model \\LQ Mass Limit(GeV) \\ for $\kappa = 0$}  & 1665 & 1785 & 1872\\
\hline
\thead{$U_1$ model \\LQ Mass Limit(GeV) \\ for $\kappa = 1$}  & 1442 & 1756 & 1831\\
\hline
\end{tabular}}
\caption{Bounds on the LQ mass for pair produced LQs decaying into up-type quarks and muon neutrinos.}
\label{tab:NeutrinoUpQuarkMuons}
\end{center}
\end{table}
\newpage
\section{Mass bounds coming from Lepton Flavour Violating processes}

In this section, we use the bounds from LFV processes presented in \cite{deMedeirosVarzielas:2015yxm} to place bounds on LQs coupling to more than one lepton flavour. The expressions show the dependence on the specific couplings for each observable, and we use these to calculate the mass bounds for the Hierarchical, Flipped, and 2-3 Democratic scenarios. Note that strictly, these bounds strictly apply to the case of scalar LQs, as the computation of the processes differs for vector mediators.

\begin{table} [H]
\begin{center}
\renewcommand\theadalign{bc}
{\renewcommand{\arraystretch}{0.8}
\begin{tabular}{|c||c|c|c|c|}
\hline
 Observable & Constraint & \thead{Hierarchical Scenario \\ LQ Mass \\ Limit (GeV)} & \thead{Flipped Scenario \\LQ Mass \\ Limit (GeV)} & \thead{2-3 Democratic Scenario \\LQ Mass \\ Limit (GeV)}\\
\hline
\hline
${\cal{B}}(\mu \to e \gamma) $ & $|  \lambda_{qe}\lambda_{q\mu}^* | \lesssim   \frac{M^2}{(34 {\rm TeV})^2} $ & $6800$ & $6800$ & $34000$ \\
${\cal{B}}(\tau \to e \gamma)$ & $|  \lambda_{qe}\lambda_{q\tau}^* |  \lesssim   \frac{M^2}{(0.6 {\rm TeV})^2} $ & $120$ & $120$ & $600$ \\
${\cal{B}}(\tau \to \mu \gamma)$ &  $|  \lambda_{q\mu}\lambda_{q\tau}^* |  \lesssim   \frac{M^2}{(0.7 \, {\rm TeV})^2}$ & $140$ & $140$ & $700$ \\
${\cal{B}}(B \to K  \mu^\pm e^\mp)$ & $\sqrt{ |  \lambda_{s \mu }\lambda_{b e}^* |^2+ |  \lambda_{b \mu }\lambda_{se}^* |^2} \lesssim   \frac{M^2}{(19.4\,  {\rm TeV})^2} $ & $4614$ & $4614$ & $23070$\\
${\cal{B}}(B \to K  \tau^\pm e^\mp)$ & $\sqrt{ |  \lambda_{s \tau }\lambda_{b e}^* |^2+ |  \lambda_{b \tau }\lambda_{se}^* |^2} \lesssim   \frac{M^2}{(3.3 \,  {\rm TeV})^2} $ & $785$ & $785$ & $3925$\\
${\cal{B}}(B \to K  \mu^\pm \tau^\mp) $ & $\sqrt{ |  \lambda_{s \mu }\lambda_{b \tau}^* |^2+ |  \lambda_{b \mu }\lambda_{s\tau}^* |^2} \lesssim   \frac{M^2}{(2.9 \,  {\rm TeV})^2}$ & $690$ & $690$ & $3450$\\
\hline
\end{tabular}}
\caption{Bounds on the LQ couplings from LFV processes ($q=d,s,b$). As in \cite{deMedeirosVarzielas:2015yxm}, we ignored tuning between leading order diagrams in the amplitudes of $\ell \to \ell^\prime  \gamma$.}
\label{tab:LFV}
\end{center}
\end{table}
\begin{figure}[H]
\begin{center}
\includegraphics[scale=0.4]{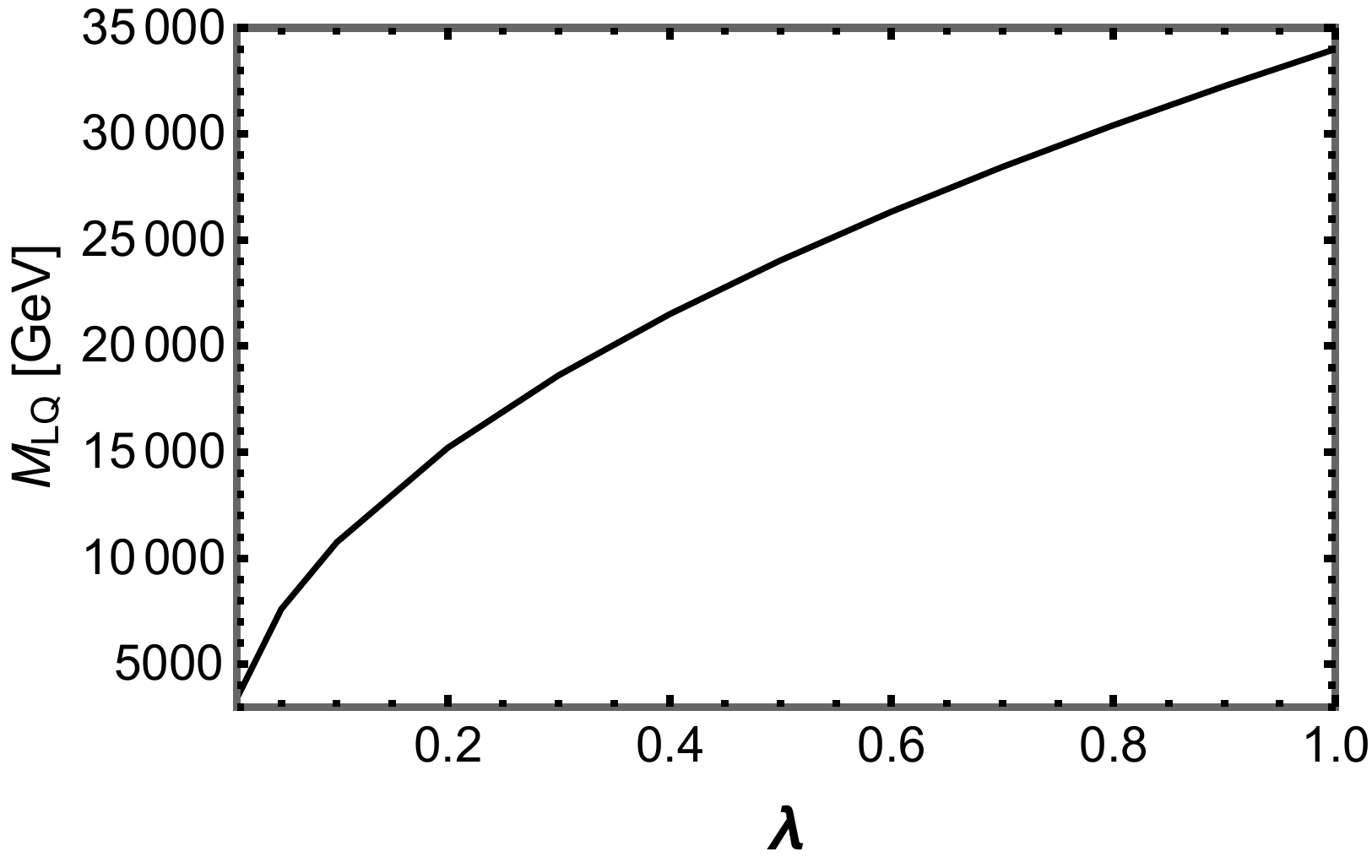}
\includegraphics[scale=0.4]{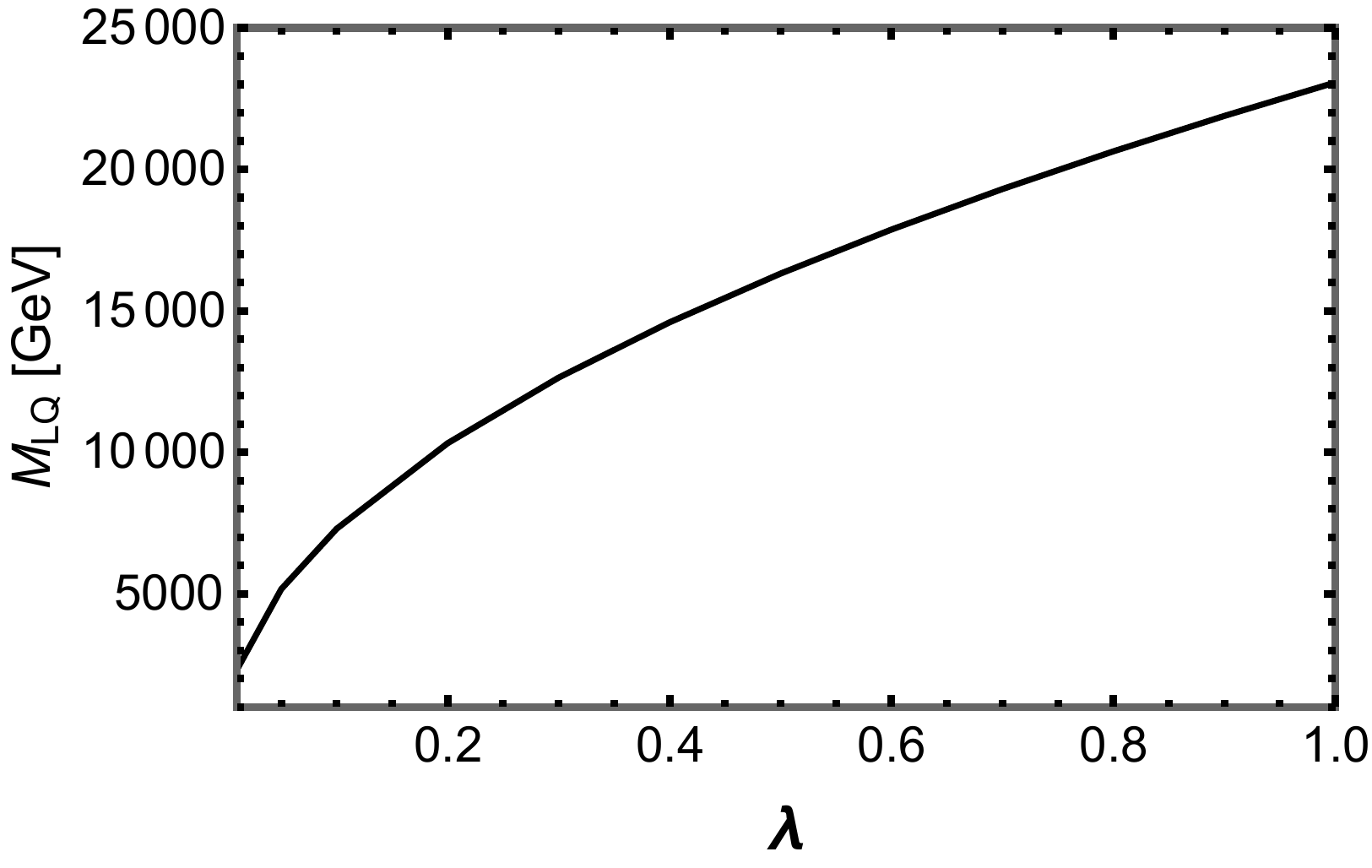}
\includegraphics[scale=0.4]{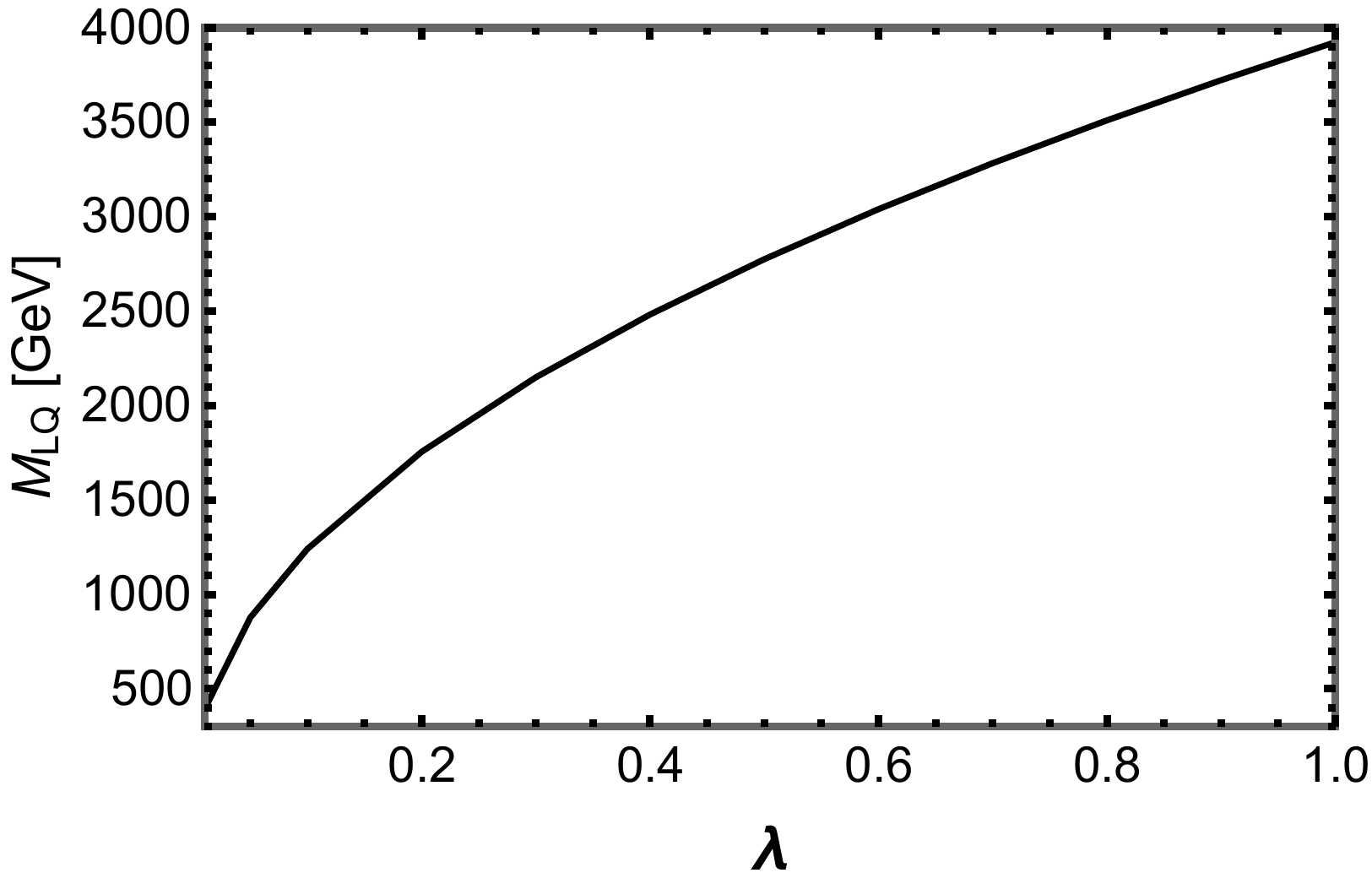}
\includegraphics[scale=0.4]{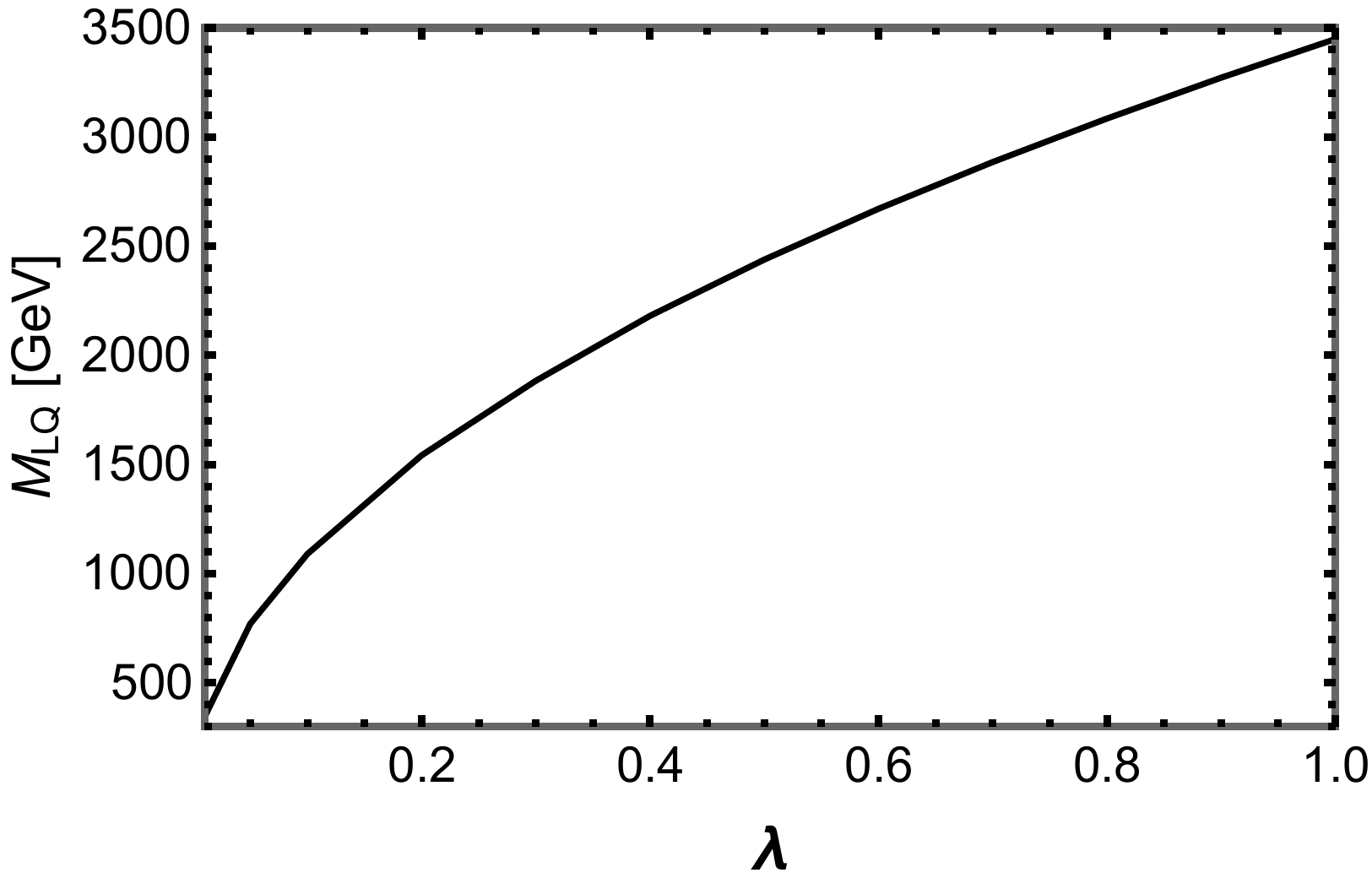}
\end{center}
\caption{Top Left- Plot for the LQ Mass (Y-axis) and LQ Coupling(X-axis) for the process $\mu \to e\,\gamma$, \, Top Right - Plot for the LQ Mass (Y-axis) and LQ Coupling(X-axis) for the process of $B \to K \,\mu\,e$ \, Bottom Left- Plot for the LQ Mass (Y-axis) and LQ Coupling(X-axis) for the process of $B \to K \,\tau\,e$ \, Bottom Right-  Plot for the LQ Mass (Y-axis) and LQ Coupling(X-axis) for the process of $B \to K \,\mu\,\tau$. In this plot we want to show how the mass limits change based on the coupling constants for the LFV constraint for different decay processes. We haven't plotted the other two channels from the table because the bounds are too low. }
\label{fig:PlotLFV}
\end{figure}
\newpage
\subsection{Summary of the results}

We now summarise the results from the previous subsections. We first note that, for scalar LQs coupling to more than one lepton flavour, we have strict bounds from Lepton Flavor Violating processes. It is clear from the constraints that these are more stringent in the 2-3 Democratic Scenario, in which case the mass of the LQ needs to be larger than 34 TeV if it couples to both of the lighter generations of charged leptons ($e \mu$). For 2-3 Democratic Scenarios coupling to $\tau$ and one of the lighter-charged leptons, the mass bound is still above 3 TeV.
The Hierarchical and Flipped Scenarios have considerably weaker bounds by about one order of magnitude, as seen in Table \ref{tab:LFV}.

With respect to constraints coming from the LHC, 2-3 Democratic Scenarios are bounded roughly at the order of 1 to 1.5 TeV (depending on whether $S_3$, $R_2$,$S_1$, $\Tilde{S}_1$, $\Tilde{R}_2$ or $U_1$ ($\kappa = 0$, $\kappa=1$). A few stronger bounds are obtained with e.g. vector LQs coupling to muons, with mass excluded below 2 TeV, and vector LQs coupling to electrons, with mass excluded below 2.5 TeV in some cases.

The constraints coming from the LHC summarised above, can be compared
to the complementary bounds arising from t-channel Drell-Yan processes, e.g. from \cite{Greljo:2022jac}, where typical mass bounds are 2.5 TeV for LFU LQ, and 4-6 TeV for Minimal Flavour Violation LQ.

With respect to the LQs $\tilde{R}_2$, $S_1$ it is interesting to note that the bounds coming from decays involving down-type quarks and leptons, the bounds are stronger than the corresponding bounds for the LQ $S_3$, which contrasts with the bounds coming from the other decays.

We note that the mass bounds obtained from decay channels involving neutrinos are always weaker than the corresponding bounds using other decay channels.


\section{Conclusions \label{sec:con}}

We have considered a Beyond the SM framework where we add either scalar or vector LQs. We classify the flavour couplings of the LQ to the SM fermions according to the leptons that do or do not couple to the LQ. These scenarios are motivated by underlying flavour symmetries that can force couplings to vanish.

For each class of LQs, we investigate the processes that lead to the strongest bounds on the LQ mass.

Apart from the expectation associated with lepton flavour violation bounds coming from $\mu$ to $e$ processes that have very high precision, the mass bounds obtained are of the order of a few TeV. The detailed bounds depend on the type of LQ, the leptons it couples to, and the specific scenario (if the couplings are 2-3 Democratic or hierarchical). We have systematically presented bounds for each of these options, providing a comprehensive approach that highlights the present constraints on flavored LQ models.

\section*{Acknowledgements}
IdMV acknowledges funding from Funda\c{c}\~{a}o para a Ci\^{e}ncia e a Tecnologia (FCT) through the contract UID/FIS/00777/2020 and was supported in part by FCT through projects CFTP-FCT Unit 777 (UID/FIS/00777/2019), PTDC/FIS-PAR/29436/2017, CERN/FIS-PAR/0004/2019, CERN/FIS-PAR/0008/2019 and CERN/FIS-PAR/0019/2021 which are partially funded through POCTI (FEDER), COMPETE, QREN and EU.
\bibliographystyle{jhep}
\bibliography{main}
\section*{Appendix}
\subsection{Leptoquark flavoured patterns \label{sec:patterns}}

We briefly review here the type of LQ couplings we are considering. We have shown the most generalized couplings for all three models and the notation should differ for different models when it is implemented in the theory and for phenomenological analysis. We note that these specific patterns can be obtained by the same type of flavour symmetries that are used to address the flavour problem of the SM \cite{deMedeirosVarzielas:2015yxm}.

We take 3 lepton isolation patterns, where the LQ couples only to electron or to muon or to tau, and 3 two-columned patterns where the LQ does not couple to electron or to muon or to tau:

\begin{equation}
\label{eq:isolation}
\lambda^{[e]}_{dl} = 
\left(
\begin{array}{ccc}
\lambda_{de} & 0 & 0  \\
\lambda_{se}  & 0 & 0    \\
\lambda_{be}  & 0  & 0 
\end{array}
\right), \,\,\,\,\,
\lambda^{[\mu]}_{dl} = 
\left(
\begin{array}{ccc}
0 & \lambda_{d\mu} & 0  \\
0 & \lambda_{s\mu} & 0    \\
0 & \lambda_{b\mu} & 0 
\end{array}
\right), \,\,\,\,\,
\lambda^{[\tau]}_{dl} = 
\left(
\begin{array}{ccc}
0 & 0 & \lambda_{d\tau}  \\
0  & 0 & \lambda_{s \tau}     \\
0  & 0  & \lambda_{b \tau}  
\end{array}
\right)   \, ,
\end{equation}
%
%
We obtain the couplings to e.g. up-type quarks by proceeding as described in \cite{deMedeirosVarzielas:2019lgb}. The coupling to up type quarks corresponding to $\lambda^{[e]}_{dl}$ (with $\lambda_{de} = 0$):
\begin{equation}
\label{eq:ue}
\lambda^{[e]}_{ul} =
\frac{1}{\sqrt{2}} \left(
\begin{array}{ccc}
V_{ub} \lambda_{be} +  V_{us} \lambda_{se} & 0 & 0 \\
V_{cb} \lambda_{be} +  V_{cs} \lambda_{se} & 0 & 0  \\
V_{tb} \lambda_{be} +  V_{ts} \lambda_{se} & 0 & 0 
\end{array}
\right) \, . 
\end{equation}
The other isolation cases have corresponding
\begin{equation}
\label{eq:umutau}
\lambda^{[\mu]}_{ul} =
\frac{1}{\sqrt{2}} \left(
\begin{array}{ccc}
0 & V_{ub} \lambda_{b\mu} +  V_{us} \lambda_{s\mu} & 0 \\
0 & V_{cb} \lambda_{b\mu} +  V_{cs} \lambda_{s\mu} & 0  \\
0 & V_{tb} \lambda_{b\mu} +  V_{ts} \lambda_{s\mu} & 0 
\end{array}
\right), \,\,\,\,\,\,\,\,\,\, 
\lambda^{[\tau]}_{ul} =
\frac{1}{\sqrt{2}} \left(
\begin{array}{ccc}
0 & 0 & V_{ub} \lambda_{b\tau} +  V_{us} \lambda_{s\tau} \\
0 & 0 & V_{cb} \lambda_{b\tau} +  V_{cs} \lambda_{s\tau}  \\
0 & 0 & V_{tb} \lambda_{b\tau} +  V_{ts} \lambda_{s\tau} 
\end{array}
\right) \, . 
\end{equation}

For the up-type quark and charged-lepton couplings $\lambda_{ul}$, rigorously the coupling matrices respective to each of the down-type couplings should be used. We use simplified structures by putting the first row to zero and checked that the difference in terms of the LQ mass bounds with respect to the rigorous structure was negligible (1 part in 10000). 
\newline For a lepton isolation case for neutrinos such as electron isolation (with $\lambda_{de} = 0$), the corresponding $\lambda_{d\nu}$ is:
\begin{equation}
\label{eq:dnu}
\lambda_{d\nu}^{[e]} =
\frac{1}{\sqrt{2}} \left(
\begin{array}{ccc}
0 & 0 & 0 \\
U_{11} \lambda_{se} & U_{12} \lambda_{se}  & U_{13}\lambda_{se}  \\
U_{11} \lambda_{be} & U_{12}\lambda_{be} & U_{13} \lambda_{be} 
\end{array}
\right)
\end{equation}
Where $U$ is the PMNS matrix.
The other isolation cases have corresponding
\begin{equation}
\label{eq:dnu_mu}
\lambda_{d\nu}^{[\mu]} =
\frac{1}{\sqrt{2}} \left(
\begin{array}{ccc}
0 & 0 & 0 \\
U_{21} \lambda_{s\mu} & U_{22} \lambda_{s\mu}  & U_{23}\lambda_{s\mu}  \\
U_{21} \lambda_{b\mu} & U_{22}\lambda_{b\mu} & U_{23} \lambda_{b\mu} 
\end{array}
\right), \,\,\,\,\,\,\,\,\,\,
\lambda_{d\nu}^{[\tau]} =
\frac{1}{\sqrt{2}} \left(
\begin{array}{ccc}
0 & 0 & 0 \\
U_{31} \lambda_{s\tau} & U_{32} \lambda_{s\tau}  & U_{33}\lambda_{s\tau}  \\
U_{31} \lambda_{b\tau} & U_{32}\lambda_{b\tau} & U_{33} \lambda_{b\tau} 
\end{array}
\right) \,.
\end{equation}
 For the couplings to up-type quarks case one obtains for $\lambda_{dl}^{[e]}$ (with $\lambda_{de} = 0$),
\begin{equation}
\label{eq:unuV1}
\lambda_{u\nu}^{[e]} =
\left(
\begin{array}{ccc}
U_{11}  \left( V^{\star}_{ub} \lambda_{be} + V^{\star}_{us} \lambda_{se} \right) & U_{12}  \left( V^{\star}_{ub} \lambda_{be} + V^{\star}_{us} \lambda_{se} \right) & U_{13} \left( V^{\star}_{ub} \lambda_{be} + V^{\star}_{us} \lambda_{se} \right) \\
U_{11}  \left( V^{\star}_{cb} \lambda_{be} + V^{\star}_{cs} \lambda_{se} \right) & U_{12} \left( V^{\star}_{cb} \lambda_{be} + V^{\star}_{cs} \lambda_{se} \right)  & U_{13} \left( V^{\star}_{cb} \lambda_{be} + V^{\star}_{cs} \lambda_{se} \right) \\
U_{11}  \left( V^{\star}_{tb} \lambda_{be} + V^{\star}_{ts} \lambda_{se} \right) & U_{12} \left( V^{\star}_{tb} \lambda_{be} + V^{\star}_{ts} \lambda_{se} \right) & U_{13} \left( V^{\star}_{tb} \lambda_{be} + V^{\star}_{ts} \lambda_{se} \right)
\end{array}
\right)\,.
\end{equation}
For $\lambda_{dl}^{[\mu]}$,
\begin{equation}
\label{eq:munuV1}
\lambda_{u\nu}^{[\mu]} =
\left(
\begin{array}{ccc}
U_{21}  \left( V^{\star}_{ub} \lambda_{b\mu} + V^{\star}_{us} \lambda_{s\mu} \right) & U_{22}  \left( V^{\star}_{ub} \lambda_{b\mu} + V^{\star}_{us} \lambda_{s\mu} \right) & U_{23} \left( V^{\star}_{ub} \lambda_{b\mu} + V^{\star}_{us} \lambda_{s\mu} \right) \\
U_{21}  \left( V^{\star}_{cb} \lambda_{b\mu} + V^{\star}_{cs} \lambda_{s\mu} \right) & U_{22} \left( V^{\star}_{cb} \lambda_{b\mu} + V^{\star}_{cs} \lambda_{s\mu} \right)  & U_{23} \left( V^{\star}_{cb} \lambda_{b\mu} + V^{\star}_{cs} \lambda_{s\mu} \right) \\
U_{21}  \left( V^{\star}_{tb} \lambda_{b\mu} + V^{\star}_{ts} \lambda_{s\mu} \right) & U_{22} \left( V^{\star}_{tb} \lambda_{b\mu} + V^{\star}_{ts} \lambda_{s\mu} \right) & U_{23} \left( V^{\star}_{tb} \lambda_{b\mu} + V^{\star}_{ts} \lambda_{s\mu} \right)
\end{array}
\right)\,.
\end{equation}
For $\lambda_{dl}^{[\tau]}$,
\begin{equation}
\label{eq:taunuV1}
\lambda_{u\nu}^{[\tau]} =
\left(
\begin{array}{ccc}
U_{31}  \left( V^{\star}_{ub} \lambda_{b\tau} + V^{\star}_{us} \lambda_{s\tau} \right) & U_{32}  \left( V^{\star}_{ub} \lambda_{b\tau} + V^{\star}_{us} \lambda_{s\tau} \right) & U_{33} \left( V^{\star}_{ub} \lambda_{b\tau} + V^{\star}_{us} \lambda_{s\tau} \right) \\
U_{31}  \left( V^{\star}_{cb} \lambda_{b\tau} + V^{\star}_{cs} \lambda_{s\tau} \right) & U_{32} \left( V^{\star}_{cb} \lambda_{b\tau} + V^{\star}_{cs} \lambda_{s\tau} \right)  & U_{33} \left( V^{\star}_{cb} \lambda_{b\tau} + V^{\star}_{cs} \lambda_{s\tau} \right) \\
U_{31}  \left( V^{\star}_{tb} \lambda_{b\tau} + V^{\star}_{ts} \lambda_{s\tau} \right) & U_{32} \left( V^{\star}_{tb} \lambda_{b\tau} + V^{\star}_{ts} \lambda_{s\tau} \right) & U_{33} \left( V^{\star}_{tb} \lambda_{b\tau} + V^{\star}_{ts} \lambda_{s\tau} \right)
\end{array}
\right)\,.
\end{equation}

\end{document}